\documentclass[10pt,journal,compsoc]{IEEEtran}
\usepackage{amsmath,amsfonts}
\usepackage{algorithmic}
\usepackage{algorithm}
\usepackage{array}
\usepackage[caption=false,font=normalsize,labelfont=sf,textfont=sf]{subfig}
\usepackage{textcomp}
\usepackage{stfloats}
\usepackage{url}
\usepackage{verbatim}
\usepackage{graphicx}
\usepackage{cite}
\usepackage{xcolor}
\usepackage[table]{xcolor} 
\usepackage{booktabs} 
\usepackage{ragged2e}
\newcolumntype{P}[1]{>{\RaggedRight\hspace{0pt}}p{#1}}
\newcolumntype{L}[1]{>{\RaggedLeft\hspace{0pt}}p{#1}}
\newcolumntype{X}[1]{>{\centering\hspace{0pt}}p{#1}}

\usepackage{hyperref}
\usepackage{enumitem}
\usepackage[many]{tcolorbox} 
\usepackage{fontawesome5}
\hypersetup{
    colorlinks=true,
    linkcolor=black,
    filecolor=black,      
    urlcolor=black,
    citecolor=black,
}

\usepackage[many]{tcolorbox} 
\definecolor{tableGray}{RGB}{243, 244, 245}
\definecolor{borderGray}{RGB}{229, 230, 233}

\newtcolorbox{boxA}{
    colback = tableGray, 
    boxrule = 0pt  
}

\def\blackbar#1#2{
   {\color{black}\rule{#1mm}{4pt}}  #2}

\newtcolorbox{boxDef}{
    colback = tableGray, 
    boxrule = 0pt,  
    colframe = borderGray,
    leftrule = 6pt,
    sharp corners
}

\begin{document}


\title{EmpathiSEr: Validated Scales for Measuring Empathy Towards Software Engineering Practitioners and Users}

\author{Hashini~Gunatilake,~\IEEEmembership{Graduate Student Member,~IEEE},
        John~Grundy~\IEEEmembership{Fellow,~IEEE},
        Rashina~Hoda~\IEEEmembership{Member,~IEEE},
        Ingo~Mueller
\IEEEcompsocitemizethanks{\IEEEcompsocthanksitem Gunatilake, Grundy, and Hoda are with Faculty of IT, Monash University, Melbourne, Australia \protect\\
Mueller is with Monash Centre for Health Research \& Implementation, Monash Health, Melbourne, Australia \protect\\
Contact E-mail: hashini.gunatilake@monash.edu}

}

\markboth{Submitted to IEEE Transactions on Software Engineering}%
{Gunatilake \MakeLowercase{\textit{et al.}}: The EmpathiSEr: Development and Validation of Software Engineering Oriented Empathy Scales}


\maketitle

\begin{abstract}
Empathy plays a critical role in software engineering (SE), \textcolor{black}{for instance in situations where developers interpret non-technical users' frustration with system usability or where product owners account for the technical constraints experienced by engineers during implementation. Such interactions shape} collaboration, communication, and user-centred design outcomes. Although SE research has increasingly recognised empathy as a key human aspect, there remains no validated instrument specifically designed to measure it within the unique socio-technical contexts of SE. Existing generic empathy scales, while well-established in psychology and healthcare, often rely on language, scenarios, and assumptions that are not meaningful or interpretable for software practitioners. These scales fail to account for the diverse, role-specific, and domain-bound expressions of empathy in SE, such as understanding a non-technical user's frustrations or another practitioner's technical constraints, which differ substantially from empathy in clinical or everyday contexts.
To address this gap, we developed and validated two domain-specific empathy scales: EmpathiSEr-P, assessing empathy among practitioners, and EmpathiSEr-U, capturing practitioner empathy towards users.
Grounded in a practitioner-informed conceptual framework, the scales encompass three dimensions of empathy: cognitive empathy, affective empathy, and empathic responses. We followed a rigorous, multi-phase methodology, including expert evaluation, cognitive interviews, and two practitioner surveys.
The resulting instruments represent the first psychometrically validated empathy scales tailored to SE, offering researchers and practitioners a tool for assessing empathy and designing empathy-enhancing interventions in software teams and user interactions. 
\end{abstract}


\section{Introduction} \label{sec:Introduction}
\IEEEPARstart{E}{mpathy} has gained increasing recognition as an important human aspect in software engineering (SE) \cite{gunatilake2024SLR, gunatilake2025theory, cerqueira2023thematic, cerqueira2024empathy, gunatilake2025manifestations}. In psychology, empathy is defined as \textit{``the ability to experience affective and cognitive states of another person, whilst maintaining a distinct self, in order to understand the other''} \cite{guthridge2021taxonomy}. As software systems are built by and for people, software practitioners regularly interact with diverse stakeholders including developers, product owners, business analysts, testers, and project managers. In these interactions, the ability to understand others' perspectives, communicate effectively, and respond thoughtfully plays a critical role in collaboration, user-centred design, and decision-making. Prior studies have linked empathy to improved team dynamics, better stakeholder engagement, and higher quality outcomes \cite{gunatilake2025theory, cerqueira2023thematic, cerqueira2024empathy}. \textcolor{black}{However, current evidence remains insufficient to determine how empathy varies across SE roles and contexts, how it relates to practitioner, team, and user outcomes, or whether initiatives intended to strengthen empathy produce meaningful changes in practice.} In the age of artificial intelligence (AI), where machines can replicate many technical tasks, empathy remains a uniquely human trait that not only defines what it means to be human, but also plays a vital role in building inclusive, ethical, and socially responsible technologies \cite{perry2023ai, rubin2025comparing}.
%
%
\textcolor{black}{Advancing this understanding requires researchers to investigate empathy systematically and cumulatively across SE settings. A reliable and context-sensitive means of assessment can support empirical comparisons across roles and teams, examination of relationships between empathy and SE outcomes, and evaluation of interventions intended to strengthen empathy-related practices \cite{graziotin2021psychometrics}. For practitioners, such assessment may also support reflection on team dynamics and inform training and process improvement initiatives \cite{gunatilake2023empathy, gunatilake2025theory, gunatilake2025manifestations, gunatilake2025guidelines}.}

Despite growing interest in empathy within SE, the field currently lacks a standardised instrument to measure empathy \cite{gunatilake2023empathy, gunatilake2024enablers, gunatilake2025theory}. While several empathy scales exist in psychology \cite{davis1980multidimensional, davis1983measuring, reniers2011qcae, baron2004empathy}, healthcare \cite{hojat2001jefferson}, and design disciplines \cite{drouet2024development}, they are not directly transferable to SE due to fundamental contextual differences in professional roles, communication norms, and collaborative practices \cite{gunatilake2023empathy, gunatilake2024enablers}.
%
\textcolor{black}{Generic empathy scales are often too broad and include items that lack contextual relevance to SE, as identified in our prior empirical work \cite{gunatilake2023empathy, gunatilake2024enablers, gunatilake2025theory, gunatilake2025manifestations}. For instance, the Interpersonal Reactivity Index (IRI) includes Fantasy and Personal Distress subscales with items items referring to reactions to fictional narratives or general social situations such as \textit{``I really get involved with the feelings of the characters in a novel,''} \textit{``I daydream and fantasize, with some regularity, about things that might happen to me,''} and \textit{``When I see someone who badly needs help in an emergency, I go to pieces,''} which are difficult to apply meaningfully in professional SE settings. Similarly, the EQ includes items that do not meaningfully reflect SE activities, such as \textit{``I prefer animals to humans''} and \textit{``When I was a child, I enjoyed cutting up worms to see what would happen.''} These scenarios lack ecological validity with respect to the professional contexts in which software practitioners exercise empathy \cite{gunatilake2023empathy}. Practitioners reported difficulty relating to such items across our prior studies \cite{gunatilake2024enablers, gunatilake2025theory, gunatilake2025manifestations}, particularly the items grounded in fictional or non-professional scenarios, and perceived them as contextually irrelevant. When practitioners are unable to meaningfully map scale items to their professional contexts, their responses may not reflect genuine empathic disposition, undermining construct validity despite the instrument's established psychometric properties in its original domain \cite{graziotin2021psychometrics}.} 
This limitation is not unique to SE. Other fields have addressed similar issues by creating domain-specific measures. The Jefferson Scale of Physician Empathy (JSPE), for example, was developed for healthcare professionals and includes items focused on doctor–patient relationships, such as \textit{``I believe that empathy is an important therapeutic factor in medical or surgical treatment''}, making it suitable for healthcare, but remaining unsuitable for SE without conceptual adaptation. 


In contrast, empathy in SE manifests through practices such as collaborating with distributed, cross-functional teams, understanding the perspectives of non-technical stakeholders, and navigating fast-paced development timelines. For example, thoughtfully responding to a user-reported bug or recognising a teammate's burnout during a sprint requires practical, context-sensitive empathy. These interactions significantly differ from emotionally intense encounters typical in clinical settings. Therefore, existing generic or domain-specific scales often fail to capture the nuanced and task-oriented nature of empathy in the socio-technical discipline \cite{hoda2022STGT} of SE.

%

\textcolor{black}{Recent SE research has begun to examine how empathy is conceptualised and enacted, drawing on practitioner-generated narratives and empirical studies \cite{cerqueira2023thematic, cerqueira2024empathy, gunatilake2025manifestations}. This work has established an initial understanding of empathy in SE, but further progress requires methods that support systematic investigation across roles, relationships, and organisational contexts. In particular, the lack of a validated SE-specific instrument limits researchers’ ability to compare findings across studies, investigate relationships between empathy and outcomes such as collaboration, communication, and user satisfaction, and evaluate whether empathy-focused interventions lead to change. A domain-specific scale can complement existing qualitative approaches by providing a consistent means of assessing empathy as it manifests in SE practice.}

\textcolor{black}{To provide this measurement foundation}, we developed and validated two SE-oriented empathy scales: \textit{EmpathiSEr-P}, for measuring empathy among software practitioners, and \textit{EmpathiSEr-U}, for measuring software practitioner empathy towards software users. 
Our methodology was informed by best practices in empathy scale development across psychology, healthcare, and design domains \cite{hojat2016jefferson, drouet2024development}. We employed a multi-phase process comprising: (1) construct definition; (2) item pool generation; (3) a pretest survey; (4) two rounds of expert evaluation; (5) cognitive interviews with practitioners; and (6) a large-scale validation survey.
The final versions of the scales were developed through iterative refinement based on conceptual alignment, practitioner and expert feedback, and psychometric analysis. 
Each scale captures cognitive, affective, and behavioural expressions of empathy. 
The key contributions of this work are as follows:
\begin{itemize}
    \item \textcolor{black}{We introduce the first psychometrically validated, domain-specific instruments for assessing self-reported manifestations of empathy in organisational and socio-technical SE contexts. These scales, \textit{EmpathiSEr-P} and \textit{EmpathiSEr-U}, capture the nuances of empathy as experienced among practitioners and in practitioner-user interactions, addressing a critical gap in SE research.}
    
    \item \textcolor{black}{We present an empirically grounded operationalisation of empathy involving three interrelated dimensions including cognitive, affective, and behavioural as they manifest in SE contexts.}
    
    \item \textcolor{black}{We present a rigorous and replicable methodology for scale development combining qualitative inquiry, expert input, and quantitative psychometric validation.} 
    
    \item \textcolor{black}{We offer key directions for future SE research and practical applications, including empathy assessment, reflective practice, and empathy-enhancing interventions in industry.}

\end{itemize}

\textcolor{black}{These scales provide instruments for assessing self-reported manifestations of empathy within organisational and socio-technical SE contexts. While psychometric validation provides evidence of reliability and construct validity of these instruments under defined conditions, it does not imply that they constitute absolute or exhaustive measures of empathy as an inherent personal attribute. Instead, the scales enable empirical analysis and reflective practice in studies of team collaboration, user-centred development, and professional training, supporting a more systematic and context-sensitive understanding of empathy in SE.}


\section{Related Work} \label{sec:Related Work}
\subsection{Conceptualisation of Empathy} \label{sec:Conceptualisation of Empathy}
Empathy is widely recognised as a multidimensional construct, encompassing \textit{cognitive}, \textit{affective (emotional)}, and \textit{behavioural} dimensions \cite{clark2019feel, cuff2016empathy}. However, there is ongoing debate regarding the nature of empathy, whether it should be viewed as predominantly cognitive, affective, or as a combination of both \cite{clark2019feel, hojat2007empathy}. Cognitive empathy refers to the ability to understand another person's thoughts, emotions, or perspective, while affective empathy involves sharing or resonating with another's emotional experience \cite{clark2019feel}. Empathy has been examined across various disciplines, including evolutionary, psychological, neuroscientific, and moral perspectives, as well as political, economic, and cultural viewpoints \cite{guthridge2020critical}. This diversity has resulted in a wide range of conceptualisations and definitions, often leading to ambiguity about the core nature of empathy. In a comprehensive content analysis of 146 leading empathy definitions, Guthridge and Giummarra identified six primary dimensions of empathy: cognitive, affective, experiential, ability-based, understanding, and self–other distinction \cite{guthridge2021taxonomy}. They proposed a definition of empathy as: ``the ability to experience affective and cognitive states of another person, whilst maintaining a distinct self, in order to understand the other.'' This definition aligns with the widely held view that empathy involves at least two key dimensions: cognition and affect \cite{decety2011neuroevolution}. \textcolor{black}{However, it does not fully capture the dimensions of empathy observed in our prior work within organisational and socio-technical SE contexts, particularly in practitioner-practitioner and practitioner-user interactions that characterise collaborative software development practice \cite{gunatilake2025manifestations}.} 
While empathy is recognised as a complex, interconnected neuropsychological process, the precise mechanisms that link its cognitive and affective components remain under-explored \cite{guthridge2023role}.

\subsection{Measures of Empathy}  \label{sec:Measures of Empathy}

A wide array of self-report empathy scales have been developed to measure empathy, differing in target population, dimensional focus, and theoretical grounding. Among the most widely used is the \textit{IRI} \cite{davis1980multidimensional, davis1983measuring}, which includes four subscales for Perspective Taking, Empathic Concern, Personal Distress, and Fantasy to capture both cognitive and affective aspects. The \textit{Empathy Quotient (EQ)} \cite{baron2004empathy}, originally developed to measure empathy in individuals with Autism, offers a single composite score that blends cognitive and emotional components. In contrast, the \textit{Questionnaire of Cognitive and Affective Empathy (QCAE)} \cite{reniers2011qcae} explicitly separates cognitive and affective empathy, allowing more granular analysis. 
Some empathy scales are tailored to specific domains. The \textit{JSPE} \cite{hojat2001jefferson, hojat2016jefferson} targets healthcare professionals, while the \textit{Consultation and Relational Empathy (CARE)} scale \cite{mercer2004consultation} focuses on empathy within clinical consultations.  The \textit{Empathy in Design (EMPA-D)} scale \cite{drouet2024development}, developed for service design domain, specifically measures service employees' empathy towards users. 

In practice, most of these instruments conceptualise empathy as a relatively stable trait rather than a dynamic state, and many blend cognitive and affective components to varying extents. Meta-analyses have identified the IRI, EQ, and QCAE as the most frequently used general-purpose measures \cite{lima2021empathy}. However, each scale presents trade-offs. The IRI and QCAE allow separate examination of cognitive and affective facets, whereas the EQ only offers an unidimensional score. General-purpose instruments like the IRI or QCAE facilitate comparisons across studies and populations but are not readily applicable for specific contexts. Further, some subscales (e.g., Personal Distress in the IRI) may conflate empathy with related constructs such as anxiety \cite{gunatilake2023empathy}. Domain-specific measures such as the JSPE or EMPA-D offer contextual relevance but may not generalise beyond their intended settings, as they are tightly tailored to the concepts and practices of their respective fields. Further, as self-report instruments, all these scales are susceptible to social desirability bias and limitations in respondents' self-awareness. 

\subsection{Empathy Research in SE} \label{sec:Empathy Research in SE} 
\textcolor{black}{Research in SE has begun to explore how empathy is manifested in the field \cite{gunatilake2025manifestations, gunatilake2025theory, cerqueira2025exploring, cerqueira2024empathy}.} Recent studies have identified key themes in practitioners' understandings of empathy, including understanding, compassion, perspective-taking, embodiment, and emotional sharing \cite{cerqueira2024empathy, cerqueira2023thematic, cerqueira2025exploring}. These themes have been organised into three broad categories: cognitive empathy, emotional empathy, and compassionate empathy. This suggests that practitioners perceive empathy as partly an intellectual capacity to understand users and team members, partly an emotional resonance with others' feelings, and partly an orientation towards care and kindness. These insights offer a conceptual foundation for understanding empathy in SE, highlighting its relevance for team collaboration, stakeholder communication, and user-centred design. However, these studies were not grounded in in-depth empirical engagement and did not provide an operational definition suitable for scale construction.
%
To address this gap, in a previous study, we conducted 22 interviews\footnote{Approved by Monash Human Research Ethics Committee. ERM Reference Number: 41060} and a large scale survey\footnote{Approved by Monash Human Research Ethics Committee. ERM Reference Number: 45708} with 116 software practitioners to empirically investigate how empathy is understood and experienced in SE contexts \cite{gunatilake2025manifestations}. Our findings revealed several dimensions of empathy that aligned with, but also extended, prior work. Specifically, we identified four key aspects of empathy in SE: \textit{cognitive empathy}, \textit{affective empathy}, \textit{empathic responses} (e.g., behavioural empathy or outward expression of empathy through actions and behaviour), and \textit{compassionate care} (e.g., concern for the well-being of others).
In another study, we developed a socio-technical grounded theory of empathy in SE based on interviews with developers and stakeholders \cite{gunatilake2025theory}. Our study demonstrated that empathy enhances software quality, fosters trust, improves communication, and helps mitigate stress and burnout among practitioners. 
In another study, we investigated the enablers and barriers to empathy in developer–user interactions and outlined practical strategies to address these challenges \cite{gunatilake2024enablers}.
Collectively, these studies emphasise the potential of empathy to enhance team dynamics, support well-being, and contribute to more effective development practices while calling for more investigations into the role of empathy in various SE practices and activities.

\subsection{Need for a Nuanced Approach to Empathy in SE} \label{sec:A Nuanced Approach to Empathy in SE}

\textcolor{black}{While empathy has received growing attention in SE research, its nuanced manifestations within SE remain under-explored \cite{gunatilake2025theory, gunatilake2024enablers}. As discussed in Section \ref{sec:Introduction}, both existing generic and domain-specific scales lack the contextual grounding and interpretability required to meaningfully assess empathy in SE contexts. Hence, there is a need for a domain-specific scale that can reliably capture empathy as it is manifested in SE contexts.} 

Software practitioners regularly interact with diverse stakeholders, each requiring different forms of empathic understanding as described in Section \ref{sec:Introduction}. 
\textcolor{black}{Studies further indicate that empathy in SE is highly situational and shaped by the characteristics of tasks, roles, and interactions \cite{gunatilake2025manifestations}. Practitioner self-reports and researcher observations have identified barriers to empathic engagement in SE, including practitioners' tendency to prioritise technical problem solving over interpersonal awareness, as well as difficulty expressing or acknowledging emotions in hierarchical or deadline driven environments \cite{gunatilake2024enablers, gunatilake2025theory, cerqueira2025exploring}.} 
\textcolor{black}{Although such barriers may also occur in other workplace settings, within SE they are shaped by technically oriented work, role dependencies, development pressures, and socio-technical interactions. Hence, empathic engagement in SE reflects not only individual tendencies but also the structural and cultural conditions under which software work is performed.}

\textcolor{black}{Many established empathy scales adopt multidimensional conceptualisations of empathy but are grounded in psychological, healthcare, or general social settings rather than SE environments. Empathic expression in SE is embedded within socio-technical workflows, role-dependent responsibilities, and task-oriented practices, and occurs across distinct relational contexts, including practitioner-user and practitioner-practitioner interactions. Existing generic empathy measures are not specifically designed to capture such contextual and relational variation in SE contexts because their items rely on language, assumptions, and scenarios that software practitioners may find difficult to relate to their work. In comparison, domains such as medicine and counselling use dedicated empathy instruments (e.g., the JSPE - health practitioner version and student version) that reflect particular professional relationships and settings \cite{lima2021empathy}.}

\textcolor{black}{Therefore, we argue for SE-specific empathy scales that integrate established conceptualisations of empathy with the socio-technical realities of SE work  \cite{gunatilake2023empathy, gunatilake2025theory}. Such scales do not assume that empathy is fundamentally different in SE; rather, they provide contextually meaningful indicators of how this broadly recognised human capacity is expressed through SE roles, relationships, and activities. They can consequently support finer-grained analysis of empathic engagement, such as understanding user needs and supporting or collaborating with peers.}

\textcolor{black}{From a measurement perspective, the choice between generic, domain-specific, and SE-specific empathy instruments depends on the level of abstraction and the purpose of analysis. Generic instruments may be appropriate for studies that require broad cross-domain comparability, while domain-specific instruments support analysis within established professional contexts such as healthcare or counselling. In contrast, SE-specific instruments are most appropriate when the research focus is on empathy as it is enacted within SE practice, where role-based interactions, task constraints, and collaborative workflows shape empathic expression. Therefore, these instruments are intended to complement, rather than replace, existing measures, enabling analysis at different levels of contextual granularity.}


\textcolor{black}{Beyond capturing contextual nuance, such scales would also enable empathy to be examined in relation to the broader landscape of factors shaping software development outcomes. Systemic and organisational challenges including resistance and failure in software implementation driven by organisational and political dynamics \cite{markus1983power}, knowledge boundaries between developers and users \cite{kensing1993pd}, and project planning and management failures, are well-established drivers of SE difficulties. Empathy, as an individual-level capability, does not replace or supersede these structural considerations. Rather, individual empathic capacity may mediate how practitioners engage with such systemic challenges: for instance, a developer's ability to take the perspective of a non-technical user may influence how effectively they bridge the knowledge gaps \cite{kensing1993pd} describes, while attunement to stakeholder experiences may shape how socio-technical misalignments are recognised and addressed in practice. However, understanding these interactions requires the ability to measure empathy in ways that are valid and meaningful within SE contexts. This gap further motivates the development of validated, SE-specific empathy scales that enable future research to examine how empathy operates alongside and potentially interacts with the organisational, methodological, and contextual factors that shape SE outcomes.}

\section{Development of SE Empathy Scales} \label{sec:Development of SE Empathy Scales}
\begin{figure*} [htbp]
    \centering
    \includegraphics[width=\textwidth]{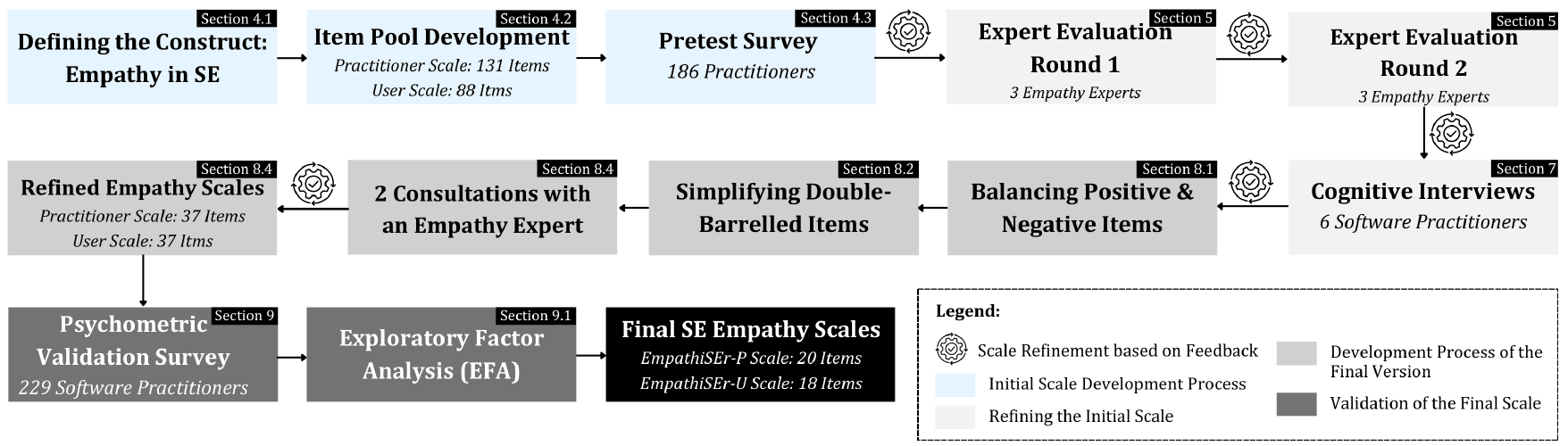}
    \caption{\textcolor{black}{Overview of the EmpathiSEr scale development and validation process. Refinement icons on the connecting arrows indicate iterative scale refinement between phases, where feedback and findings from each phase informed modifications prior to the subsequent phase.}}
    \label{fig:Overview of the SE Empathy Scale Development Process}
\end{figure*}

We followed the scale development procedures outlined in the Jefferson Scale of Empathy \cite{hojat2001jefferson, hojat2016jefferson} and the EMPA-D scale \cite{drouet2024development, cabrera2010author}. The development of the SE empathy scales followed a multi-phase, mixed-methods approach, grounded in both theoretical foundations and empirical insights. 
\textcolor{black}{The SE-specificity of the proposed instruments lies in the operationalisation of empathy through SE relevant interactional contexts and socio-technical role structures, rather than in lexical differences in item wording.}
This process (Figure \ref{fig:Overview of the SE Empathy Scale Development Process}) comprised seven main stages, with steps 1-3 forming the initial scale development phase:

\begin{enumerate}
    \item \textbf{Defining the Construct:} A literature review and qualitative studies (interviews and surveys) to explore how empathy is understood and manifested within SE contexts.

    \item \textbf{Item Pool Development:} Identified candidate items from existing empathy scales and developed new items to capture SE-specific expressions of empathy. We applied clear inclusion and exclusion criteria and mapped the items to empathy constructs relevant to SE.

    \item \textbf{Pretest Survey} (n = 186): A survey to assess the clarity and relevance of the generated item pool. The findings informed the refinement of items and the development of the initial version of the scale.

    \item \textbf{Expert Evaluation:} Interviewed empathy experts to review the items, assessing whether they accurately measured empathy and were appropriately categorised (e.g., cognitive, affective, behavioural).

    \item \textbf{Cognitive Interviews:} Conducted think-aloud interviews with software practitioners to refine item wording and ensure items were interpreted as intended.

    \item \textbf{Developing the Final Scales}: We balanced positively and negatively worded items, simplified double-barrelled items, and conducted two consultation sessions with an empathy expert to finalise the scales.

    \item \textbf{Psychometric Validation Survey} (n = 229): A large-scale survey and performed psychometric analyses to evaluate the reliability, validity, and factor structure of the final empathy scale.

\end{enumerate}

\textcolor{black}{The phases outlined above are described in detail in Sections \ref{sec: Initial Scale Development Process}, \ref{sec:Expert Evaluation}, \ref{sec:Face and Content Validity}, \ref{sec:Cognitive Interviews}, \ref{sec:Final Version of the SE-Oriented Empathy Scales}, and \ref{sec:Psychometric Analyses of the SE-Oriented Empathy Scales} which document the development, refinement, and validation of the proposed scales.}
The final versions of the EmpathiSEr scales are presented in Table \ref{tab:Final EmpathiSEr-P Scale} and \ref{tab:Final EmpathiSEr-U Scale}.

\begin{table} [htbp]
    \scriptsize
    \centering
    \caption{EmpathiSEr-P Scale (20-Item) for measuring empathy among software practitioners}
    \label{tab:Final EmpathiSEr-P Scale}
     \resizebox{\columnwidth}{!}{
    \begin{tabular}{P{0.04\linewidth} P{0.9\linewidth}}
    \toprule
    1. & I make an effort to understand the work-related needs and concerns of co-workers, by considering how things look from their perspective.\\
    2. & I find it difficult to remain flexible when co-workers face personal emergencies, especially if it affects scheduled tasks. \\
    3. & I feel uneasy during conflicts, disagreements, or tension among co-workers.\\
    4. & I tend to give feedback without considering how it might be perceived from my co-workers' perspective.\\
    5. & I try to understand the actions and emotions of co-workers.\\
    6. & During my interactions with co-workers, I am emotionally affected due to their emotional tension, stress, and frustration.\\
    7. & I do not actively consider co-workers' perspectives or work-related needs when developing software solutions.\\
    8. & When co-workers point out issues in my work, I usually focus on defending my choices rather than understanding their perspective.\\
    9. & I take co-workers' perspectives into account when discussing different viewpoints during meetings, such as code reviews or design sessions.\\
    10. & I tend to provide feedback without adjusting my tone or approach based on how it might affect co-workers.\\
    11. & I am emotionally sensitive to the emotional states of co-workers during high-pressure project situations, such as tight deadlines or significant challenges.\\
    12. & When working in a team, I don't always make an effort to ensure everyone feels heard and understood.\\
    13. & I understand the technical and domain-specific challenges co-workers face by considering their context, including their individual strengths and weaknesses.\\
    14. & I tend to make decisions without fully considering all relevant perspectives or all sides of a disagreement.\\
    15. & I try to understand the concerns of co-workers about proposed solutions by taking their perspective into account.\\
    16. & When assigning work, I do not consider co-workers' expertise, capacity, or capability.\\
    17. & I pay attention to co-workers' work styles and communication preferences to better understand their perspective.\\
    18. & I don't adjust my communication style based on co-workers' roles or level of technical expertise.\\
    19. & I clarify co-workers' doubts about requirements or domain knowledge by taking their perspective into account.\\
    20. & I try to understand ethnic and cultural differences that influence the communication and behaviour of co-workers.\\
    \bottomrule
    \end{tabular}
     }
    
\end{table}

\begin{table} [htbp]
    \scriptsize
    \centering
    \caption{EmpathiSEr-U Scale (18-Item) for measuring software practitioner empathy towards software users}

    \label{tab:Final EmpathiSEr-U Scale}
    \resizebox{\linewidth}{!}{
    \begin{tabular}{P{0.04\linewidth} P{0.9\linewidth}}
    \toprule
    1. & I make an effort to understand the needs and concerns of users, by considering how things look from their perspective.\\
    2. & I get emotionally affected when users express negative emotions due to issues in our software. \\
    3. & I don't usually take users' experiences or emotional states into account when responding to them.\\
    4. & I consider users' needs to guide development decisions, rather than focusing solely on completing tasks.\\
    5. & I don't usually consider the users' perspective to understand their experience of using the software.\\
    6. & I often become emotionally engaged with users' feelings during conflicts or disagreements.\\
    7. & I don't make an effort to build a human-level connection with users.\\
    8. & I try to understand the concerns of users about our solutions by considering their perspective and feedback.\\
    9. & I am emotionally affected when users face challenges while using the software we developed.\\
    10. & Generally I am able to make decisions based on my knowledge and without being overly influenced by emotions of users.\\
    11. & I clarify users' doubts about the software by taking their perspective into account.\\
    12. & I am emotionally moved when users share personal stories about how our software has helped them.\\
    13. & I rarely reflect on my interactions with users to understand their emotions and perspectives.\\
    14. & I consider users' needs and perspective when creating documentation to ensure it is easy to understand.\\
    15. & I get emotionally involved with the users' experiences and emotional responses of using our software.\\
    16. & I do not make an effort to understand the technical limitations and other challenges users face from their perspective.\\
    17. & I get deeply involved with the feelings of users during failures in our software.\\
    18. & When working with users, I don't always try to understand how things look from their perspective.\\
    \bottomrule
    \end{tabular}
    }
    
\end{table}

\section{Initial Scale Development Process} \label{sec: Initial Scale Development Process}

\subsection{Definition and Scope of the Construct} \label{sec:Definition and Scope of the Construct}
The first step in developing the empathy scales for SE was to define the construct in a way that captured its \textcolor{black}{contextually salient manifestations} in SE practice. While empathy is commonly understood as the ability to understand and share the feelings and perspectives of others, its operationalisation within software teams requires a more nuanced and context-specific interpretation.
To understand how empathy has been conceptualised in SE, we conducted a review of relevant literature (Section \ref{sec:Conceptualisation of Empathy}, \ref{sec:Empathy Research in SE}). 
\textcolor{black}{We identified cognitive empathy, affective empathy, empathic responses, and compassionate care as key dimensions through which empathy has been conceptualised and discussed within the SE literature \cite{gunatilake2025manifestations, cerqueira2023thematic, cerqueira2024empathy}. These dimensions are not unique to SE. However, prior empirical studies indicate that they are manifested through the socio-technical interactions, role relationships, and work practices characteristic of SE contexts.}

For the purposes of this scale, we focused on cognitive and affective empathy, as these emerged as the most prominent and foundational categories in practitioner accounts. While empathic responses were frequently mentioned, they were treated as outcomes or manifestations of cognitive and affective empathy rather than core dimensions of the construct itself. 
Similarly, compassionate care, while recognised as important, was considered beyond the scope of this work, and is already addressed by dedicated compassion scales in other domains \cite{neff2022self, raes2011construction}.
As no commonly accepted definition of empathy existed in SE, 
we adopted the meta-definition proposed by Guthridge and Giummarra (Section \ref{sec:Conceptualisation of Empathy}) \cite{guthridge2021taxonomy}. 
\textcolor{black}{This definition provided a comprehensive, domain-independent foundation that captured the multidimensional nature of empathy. It was adopted as the conceptual basis for this work because it could be operationalised within the practitioner-practitioner and practitioner-user interactional contexts identified in prior SE studies \cite{gunatilake2025manifestations, cerqueira2023thematic, cerqueira2024empathy}.}
Together, this definition and our practitioner-informed dimensions formed the conceptual basis for the development of our SE-oriented empathy scales.

\subsection{Item Pool Generation} \label{sec:Item Pool Generation}
The item pool refers to the initial set of candidate statements developed to comprehensively capture different aspects of empathy before refining them into the final scale. To develop this, we followed a two-step approach: (1) reviewing established empathy scales from various domains, and (2) incorporating insights from our prior empirical study with software practitioners \cite{gunatilake2025manifestations}. This approach enabled us to identify both theoretically grounded items and SE-specific expressions of empathy. We developed two separate item pools: one focusing on empathy between practitioners, and another on practitioner empathy towards users.
We examined several leading empathy instruments, including the 
IRI, JSPE, QCAE, EQ, and EMPA-D. Items from these scales were screened using clearly defined inclusion and exclusion criteria. Items were included if they aligned with our conceptualisation of empathy in SE and could be meaningfully interpreted within the SE context. Items were excluded if they were too abstract (\textit{e.g., ``If anyone asked me if I liked their haircut, I would reply truthfully, even if I didn't like it''}), clinically oriented (\textit{e.g.,``My understanding of how my patients and their families feel does not influence medical or surgical treatment''}), emotionally ambiguous (\textit{e.g., ``I find it hard to know what to do in a social situation''}), or unlikely to occur in professional SE environments (\textit{e.g., ``I really get involved with the feelings of the characters in a novel''}).
To address gaps in contextual relevance, we also wrote new items grounded in the experiences reported by software practitioners \cite{gunatilake2025theory, gunatilake2025manifestations}. These captured empathy-related practices such as understanding stakeholder or user needs and constraints, fostering positive team dynamics, bridging technical and non-technical perspectives, and reflect consideration for teammates' perspectives. 
The final item pool comprised 131 items addressing empathy among practitioners and 88 items focused on empathy towards users, providing comprehensive coverage of the construct in both interpersonal and user-facing dimensions of SE practice.

\subsection{Initial Version and Pretest} \label{sec:Initial Version and Pretest}
Based on the item pool, we created initial versions of two scales: one measuring empathy among practitioners and the other measuring practitioner empathy towards users. Each item was framed as a statement reflecting an empathy-related behaviour, perception, or disposition.

To evaluate the initial item set, we conducted a pretest survey\footnote{Approved by Monash Human Research Ethics Committee. ERM Reference Number: 45707} with 186 software practitioners, recruited via Prolific\footnote{https://www.prolific.com/}. \textcolor{black}{Participants received monetary compensation at a rate of approximately 7.60 GBP per hour, consistent with the platform's recommended payment guidelines at the time of data collection.}
Prolific was chosen for its ability to provide access to a diverse, pre-screened participant pool and its support for precise sampling control. Unlike recruitment through professional networks such as LinkedIn or personal contacts, which may introduce bias by restricting the sample to specific communities or social circles, Prolific enabled access to participants from a broad range of regions, industries, and experience levels. It also applies rigorous participant screening and verification mechanisms, thereby reducing the risk of low quality or fraudulent responses.
We applied specific screening criteria on Prolific to ensure the selection of relevant participants for our study. Participants were required to work in the information technology sector, have at least three years of professional experience, and regularly interact with colleagues. We applied iterative, purposive sampling to improve geographical and gender diversity, adjusting selection criteria across recruitment batches. To maintain data quality, we included attention-check questions to verify participant engagement and made all open-ended questions mandatory to encourage thoughtful, reflective responses. We excluded all responses that failed the attention-check questions and re-recruited participants to maintain data quality. Table \ref{tab:Demographic Information of the Pretest Survey Practitioners} provides a summary of the demographic and team related information of these participants.

During the survey, participants were asked to rate the relevance of each item for measuring empathy in SE contexts. In addition, they were invited to suggest modifications to improve item clarity, ensure ease of understanding, and eliminate ambiguity. They were also asked to propose any additional items they considered important for inclusion in a scale intended to measure empathy in SE. The pretest served multiple purposes: it helped identify items that were unclear, confusing, or contextually irrelevant; established which items were perceived as most representative of empathy in SE; and informed our initial decisions regarding item retention, refinement, and categorisation.

Based on the feedback, we systematically revised the item pool. First, we excluded any item where 20\% or more of the participants rated it as irrelevant, setting an inclusion threshold of 80\% agreement. We then reviewed the wording-related suggestions from participants. If a previously excluded item received multiple substantive suggestions for improvement, we revised and re-included it. We incorporated all recommended modifications that clarified or simplified the wording of existing items. In parallel, we reviewed the new items suggested by participants and included those that aligned with our conceptual framework.
Following these updates, we reviewed the revised scales for redundancy and removed items that were semantically similar or overlapping. Finally, we conducted a holistic review of item phrasing, incorporating practitioner suggestions to ensure clarity, consistency, and contextual appropriateness.
The resulting modified scales comprised 82 items focused on empathy among practitioners and 46 items measuring practitioner empathy towards users.

\begin{table*}[htbp]
    \caption{Overview of the Demographics and Team Contexts of Pretest Survey Participants}
    \label{tab:Demographic Information of the Pretest Survey Practitioners}
    \resizebox{\textwidth}{!}{
    \begin{tabular}{@{}llllll@{}}
        \toprule
        \textbf{Role} & \textbf{\# of Practitioners} & \textbf{Age} & \textbf{\# of Practitioners} &  \textbf{Experience} & \textbf{\# of Practitioners} \\
        \midrule
        
        Developer & \blackbar{48.5}{97} & 20-30 & \blackbar{27}{54} & 3-5 years & \blackbar{31}{62}\\
        
        Stakeholder & \blackbar{44.5}{89} & 31-40 & \blackbar{34}{68} & 5-10 years & \blackbar{31}{62} \\

        & & 41-50 & \blackbar{16.5}{33} & 10-15 years & \blackbar{14.5}{29}\\
        
        \cmidrule(r){1-2} 
        
        \textbf{Gender} & \textbf{\# of Practitioners} & 51-60 & \blackbar{12.5}{25} & 15-20 years & \blackbar{7.5}{15}\\

        \cmidrule(r){1-2} 
        
        Female & \blackbar{32.5}{65} & 61-70 & \blackbar{2}{4} & 20-30 years & \blackbar{7.5}{15}\\
        Male & \blackbar{60}{120} & Above 70 & \blackbar{1}{2} & 30-40 years & \blackbar{1}{2} \\
        Prefer not to say & \blackbar{0.5}{1} & Prefer not to say & \blackbar{0.5}{1} & 40-50 years & \blackbar{0.5}{1}\\
        
        \cmidrule(r){1-6} 

        \textbf{Team Size} & \textbf{\# of Practitioners} & \textbf{Org. Size} & \textbf{\# of Practitioners} & \textbf{Development Method Used} & \textbf{\# of Practitioners}\\
        
        \cmidrule(r){1-6} 

         No members & \blackbar{2.5}{5} & Self-Employed & \blackbar{3}{6} & Traditional (Waterfall) & \blackbar{49.5}{99}\\

         Less than or equal to 5 & \blackbar{18}{36} & Startup & \blackbar{5}{10} & Agile - Scrum & \blackbar{60.5}{121}\\

         5 - 10 & \blackbar{41.5}{83} & Small & \blackbar{16}{32} & Agile - Kanban & \blackbar{30}{60}\\

         10 -20 & \blackbar{20}{40} & Medium & \blackbar{25}{50} & XP & \blackbar{16.5}{33}\\

         More than 20 & \blackbar{11}{22} & Large & \blackbar{44}{88} &  & \\

         \cmidrule(r){1-6} 
         \textbf{Country} & \textbf{\# of Practitioners} & \textbf{Country} & \textbf{\# of Practitioners} & \textbf{Country} & \textbf{\# of Practitioners}  \\
         \cmidrule(r){1-6} 

         USA & \blackbar{31}{62} & Chile & \blackbar{3}{6} & Viet Nam & \blackbar{0.5}{1} \\
         UK & \blackbar{14.5}{29} & Australia & \blackbar{3}{6} & Netherlands & \blackbar{0.5}{1}  \\
         Italy & \blackbar{6}{12} & Germany & \blackbar{2.5}{5} & Brazil & \blackbar{0.5}{1}  \\
         Canada & \blackbar{5.5}{11} & Spain & \blackbar{2}{4} & Czech Republic & \blackbar{0.5}{1}  \\
         Portugal & \blackbar{4}{8} & Hungary & \blackbar{2}{4} & Croatia & \blackbar{0.5}{1}  \\
         Greece & \blackbar{4}{8} & France & \blackbar{2}{4} & Slovakia & \blackbar{0.5}{1}  \\
         Poland & \blackbar{3.5}{7} & Mexico & \blackbar{1.5}{3} & Ireland & \blackbar{0.5}{1} \\
         India & \blackbar{3.5}{7} & New Zealand & \blackbar{1}{2} & Estonia & \blackbar{0.5}{1} \\
            
        \bottomrule
    \end{tabular}
    }

\end{table*}

\begin{table*} [htbp]
    \centering
    \footnotesize
    \setlength{\aboverulesep}{0pt}
    \setlength{\belowrulesep}{0pt}
    \setlength{\extrarowheight}{.5ex}
    \caption{Overview of the Demographic and Background Information of Expert Evaluation Participants}
    \label{tab:Overview of the Demographic Information and Backgrounds of Expert Evaluation Participants}
    \resizebox{\textwidth}{!}{
    \begin{tabular}{P{0.02\textwidth} P{0.05\textwidth} P{0.05\textwidth} P{0.06\textwidth} P{0.1\textwidth} P{0.05\textwidth} P{0.22\textwidth} P{0.53\textwidth}}
        \toprule
        \textbf{ID} & \textbf{Country} & \textbf{Age Group} & \textbf{Gender} & \textbf{Professional Role} & \textbf{Experience} & \textbf{Educational Background} & \textbf{Professional Experience in Empathy or Psychology}\\
        \midrule

        \rowcolor{tableGray}
        E1 & New Zealand & 51-60 & Male & Empathy Research Consultant, Educationalist, Academic & 30+ Years & Post Graduate Diploma in Psychology, Master of Educational Psychology, and Doctor of Education. & Postdoctoral research in Education, with a thesis on the nature and function of empathy in synchronous multimedia conferencing; subsequent postgraduate research on empathy through a bicultural Maori lens. Contributed a chapter to The Psychology of Empathy and co-authored the book Mindful Empathy, focusing on enhancing empathic communication through mindfulness. Conducted consultancy work in education to support the development of empathy, and engaged in facilitation of empathy circles.\\
        
        E2 & Brazil & 31-40 & Female & Empathy Researcher & 4 Years & Bachelor's in Computer Engineering and a Master's in Computer Science, focusing on SE, and currently researching empathy in SE during PhD. & Educational and professional background in SE with a research focus on human aspects in SE. Current research investigates the role of empathy in SE, with particular interest in its impact on software quality, communication, collaboration, and related development activities.\\

        \rowcolor{tableGray}
        E3 & Australia & 41-50 & Female &  Empathy Trainer/ Speaker & 9 Years & Bachelor's degree in Exercise Science, including coursework in sports and human factors psychology. Postgraduate certificate in Nonprofit and Philanthropy, with a strong emphasis on human-centred design and social impact. & Co-developed a Virtual Reality experience simulating a psychotic episode, accompanied by an empathy workshop designed in collaboration with a psychologist. Delivered the program to professionals working with psychosis (e.g., psychologists, teachers, police) between 2016 and 2020. In response to the pandemic, founded a company related to Empathy in 2020 to translate insights from the VR project into evidence-based content and educational materials aimed at general audiences. Also lectures on Social Change at a university.\\
        
        E4 & USA & 41-50  & Female & Internal Communications \& Change Management Consultant & 20+ Years & Bachelor's degree in Marketing, with coursework emphasising empathy-related themes across subjects such as organisational psychology, consumer behaviour, and product development. & Worked as an independent consultant, helping organisations enhance customer engagement and business outcomes through empathy and effective communication. Became co-founder and CEO of a software company specialising in legacy code maintenance and modernisation. Established empathy as a foundational principle across culture, operations, and technical practices. Introduced and published the concept of ``empathy-driven development,'' which gained international recognition through keynote invitations and features in prominent industry publications. \\

        \rowcolor{tableGray}
        E5 & Australia & 31-40 & Male & Postdoc Researcher in Psychology & 13 Years & Undergraduate, honours, \& PhD studies in psychology. Honours research focused on emotion regulation, while PhD research examined social psychology of information sharing and withholding. & Completed a four-year postdoctoral position focused on human-centred AI, applying theories of basic psychological needs and social identity to examine the impact of AI systems and other technologies on users. Additional research experience includes topics in social psychology such as climate anxiety, trust, intelligence, behaviour change, and digital identity.\\
        
        E6 & Serbia & 20 - 30 & Female & Researcher & 4 Years & A specialist (6 years of edu) in Clinical Psychology, trained in cognitive-behavioural therapy (CBT) and acceptance and commitment therapy (ACT). & Worked as CBT consultant for 2 years, and currently working as a human-computer interaction researcher for 4 years.\\

         \bottomrule
    \end{tabular}}
    
\end{table*}

\section{Expert Evaluation} \label{sec:Expert Evaluation}

\textcolor{black}{Following the pretest, we conducted an expert evaluation\footnote{Approved by Monash Human Research Ethics Committee. ERM Reference Number: 46046} to assess the conceptual validity of the item pool as measures of empathy, specifically whether items accurately reflected empathy constructs and were appropriately categorised into cognitive and affective dimensions.}
We invited six experts with backgrounds in empathy research, psychology, and SE. Experts were selected based on their experience with empathy-related research or applied work in psychology, with some specialising in practice-based empathy roles (e.g., empathy coaches). Table \ref{tab:Overview of the Demographic Information and Backgrounds of Expert Evaluation Participants} provides an overview of their demographic and disciplinary backgrounds. Experts were recruited via LinkedIn and personal networks.

The expert evaluation was conducted in two rounds, each involving three experts, two with academic backgrounds in empathy or psychology, and one with a more practice-oriented perspective. This composition allowed us to balance theoretical rigour with practical relevance. Each round comprised two stages: an online survey followed by a semi-structured interview. In the survey, experts were asked to assess whether the items accurately measured empathy, evaluate their alignment with the defined dimensions of empathy (cognitive and affective), identify any redundancy or gaps in coverage, comment on the clarity and interpretability of item wording, and evaluate the appropriateness of the response scale in capturing the nuances of empathic behaviour within SE contexts.
Follow-up interviews were then conducted to clarify and elaborate on expert responses. These interviews focused on understanding why certain items were identified as not measuring empathy or flagged as duplicates, and how experts interpreted and categorised items as cognitive or affective empathy.

\textbf{In the first round of evaluation}, expert feedback led to several substantive refinements. Items that were identified as not reflecting empathy or lacking direct relevance were removed, and duplicate items were also eliminated. The remaining items were classified into cognitive and affective empathy based on the responses provided by the experts. In cases of conflicting classifications, we referred to the rationale provided during interviews and prioritised arguments that were conceptually sound and aligned with our working definition of empathy, while also considering majority agreement among the experts. Experts also frequently highlighted instances of double-barrelled wording and recommended simplifying such items to improve clarity and ensure that each item measured a single, distinct concept. In addition, experts noted that several items were more aligned with empathic responses or behavioural expressions of empathy, rather than internal empathic experience. Although we excluded empathic responses as a dimension of empathy during our initial conceptual framework (Section \ref{sec:Definition and Scope of the Construct}), all experts strongly recommended its inclusion as a distinct and meaningful component within the SE context. Based on this consistent feedback, we revised our decision and incorporated empathic response items as a third dimension of the scale. In addition, we introduced several new items based on expert suggestions and reworded or merged existing items to enhance clarity and reduce redundancy. The resulting revised scales comprised 34 items measuring empathy among practitioners and 35 items measuring practitioner empathy towards users.

\textbf{In the second round}, experts reviewed the revised scales using the same procedures. This time, they were asked to classify each item into one of the three dimensions: cognitive empathy, affective empathy, or empathic responses. Follow-up interviews again focused on the rationale for their classifications and feedback on item clarity. Based on this round of evaluation, we further refined the scales by removing additional duplicates, excluding any items not deemed to measure empathy, and adjusting wording to enhance clarity and reduce ambiguity. Additional items were added where experts identified conceptual gaps. The final output of this phase comprised 38 items addressing empathy among practitioners and 38 items measuring practitioner empathy towards users.


\section{Establishing Face and Content Validity} \label{sec:Face and Content Validity}
Establishing the validity of a scale begins with ensuring that its items are perceived as relevant and representative of the construct being measured. In this study, both face validity and content validity were addressed during the early stages of item development and refinement.

\textbf{Face validity} refers to the extent to which the items appear, on the surface, to measure the intended construct. It is typically assessed through subjective judgment, often by individuals who are not experts in the domain \cite{hojat2016jefferson}. In our case, face validity was established through the initial construction of the item pool by the research team (Section \ref{sec:Item Pool Generation}), who drew on theoretical literature and practitioner insights to ensure that each item reflected \textcolor{black}{manifestations of empathy identified in prior SE literature and practitioner studies \cite{gunatilake2025manifestations, gunatilake2024enablers, gunatilake2025theory}.}

\textbf{Content validity} involves a more systematic evaluation of a scale's contents to confirm the relevance, coverage, and representativeness of the items in relation to the construct being measured. It is usually assessed by domain experts who examine whether the scale comprehensively represents the behaviours or attributes of interest \cite{anastasi1976psychological, hojat2016jefferson}. We ensured content validity through both the pretest survey with 186 experienced software practitioners (Section \ref{sec:Initial Version and Pretest}) and expert evaluation (Section \ref{sec:Expert Evaluation}). 
\textcolor{black}{The practitioner survey helped confirm the contextual relevance, clarity, and coverage of the items, while expert feedback ensured alignment with the defined dimensions of empathy and identified conceptual gaps or redundancies in the scale. Together, these processes helped establish that the scales adequately captured a range of behaviours and expressions of empathy identified in prior SE literature and practitioner accounts \cite{gunatilake2025manifestations, gunatilake2024enablers, gunatilake2025theory}.}
\textcolor{black}{These behaviours include, for example, understanding and responding to non-technical user feedback during system development, recognising and accommodating constraints faced by teammates during implementation, and adjusting communication to align with the technical or non-technical background of different stakeholders. Items in the scale were designed to collectively capture such recurring manifestations of empathy identified across SE studies and practitioner accounts.}
\textcolor{black}{Together, these processes helped ensure that the scale items were both intuitively appropriate (face validity) and reflected a broad range of empathy manifestations identified in SE literature and practitioner accounts (content validity).}


\section{Cognitive Interviews} \label{sec:Cognitive Interviews}
To further ensure that the scales were accessible, interpretable, and meaningful to software practitioners, we conducted cognitive interviews\footnote{Approved by Monash Human Research Ethics Committee. ERM Reference Number: 46046} with six software practitioners. Cognitive interviewing is a widely used method for exploring how respondents interpret and respond to survey items, particularly when responses require conscious processing and reflection rather than automatic judgement \cite{castillo2013cognitive}. The technique is especially valuable when participants are expected to recall experiences, evaluate behaviours, or make nuanced distinctions, allowing researchers to probe the mental processes underlying their responses. When such reflective processing is absent, cognitive interviewing yields limited insights.
\textcolor{black}{The objective of this phase was to examine how software practitioners interpreted, understood, and responded to the proposed scale items in relation to their experiences in SE practice.}
\textcolor{black}{This phase involved a small, purposively selected sample, consistent with its role in identifying item interpretation rather than establishing statistical or cross-cultural generalisability.}
Participants were recruited through LinkedIn and personal networks and represented a variety of roles in the software industry, including developers, UX designers, testers, and project managers. We aimed to ensure diversity in professional background, age, and years of experience. Prior to the interviews, participants completed a pre-interview questionnaire to collect demographic information, summarised in Table \ref{tab:Overview of the Demographic Information of Cognitive Interview Participants}. During the interviews, participants were asked to think aloud as they responded to each item in the scale. For each item, they were prompted to explain how they interpreted the statement, whether it was easy to understand, and whether it related to their real-world work experience. They also provided feedback on confusing or ambiguous wording, suggested possible improvements, commented on the appropriateness of response options, and rated themselves on the item while verbalising the reasoning behind their chosen rating.

The cognitive interviews revealed several subtle but important areas for improvement. Based on this feedback, we made targeted revisions to improve clarity, simplify language, and enhance contextual adaptability. One commonly raised issue was the phrase \textit{``imagining myself in their shoes,''} which participants found difficult to relate to. While they emphasised the importance of understanding others' challenges, they preferred a more objective approach focused on recognising others' perspectives, strengths, and limitations, rather than emotionally placing themselves in another's position. 
The phrase was also perceived as vague and open to multiple interpretations. Participants suggested clearer alternatives such as \textit{``understanding their perspective''} or \textit{``understanding their technical limitations and challenges by recognising their strengths and weaknesses,''} which were seen as more reflective of their experience.
Other wording issues also emerged. For instance, participants noted that the phrase \textit{``heavy workload''} could be interpreted in multiple ways. One suggestion was to reframe it as \textit{``excessive work''} that may exceed a person's skills or capacity, rather than simply indicating a high volume of tasks. Items that a majority of participants found difficult to relate to or interpret clearly were removed. Several practitioners identified double-barrelled items as problematic, noting that such statements made it difficult to provide accurate ratings when they agreed with one part of the item but not the other. Based on this feedback, we also revised these items to ensure each statement addressed only one idea.
The final outcome of this phase was revised scales comprising 37 items focused on empathy among practitioners and 37 items measuring practitioner empathy towards users.

\begin{table*} [htbp]
    \centering
    \footnotesize
    \setlength{\aboverulesep}{0pt}
    \setlength{\belowrulesep}{0pt}
    \setlength{\extrarowheight}{.5ex}
    \caption{Overview of the Demographic Information of Cognitive Interview Participants}
    \label{tab:Overview of the Demographic Information of Cognitive Interview Participants}
    \resizebox{\textwidth}{!}{
    \begin{tabular}{P{0.1\textwidth} P{0.1\textwidth} P{0.1\textwidth} P{0.1\textwidth} P{0.2\textwidth} P{0.2\textwidth} P{0.2\textwidth}}
        \toprule
        \textbf{ID} & \textbf{Country} & \textbf{Age Group} & \textbf{Gender} & \textbf{Professional Role} & \textbf{Years of Experience} & \textbf{Domains}\\
        \midrule

        \rowcolor{tableGray}
        P1 & Canada & Above 70 & Male & Software Architect & More than 50 years & Telecommunication, Cyber-Physical Systems \\
        
        P2 & USA & 51-60 & Male & Software Quality Director & Between 20-30 years & Field Service Management, Insurance\\
        
        \rowcolor{tableGray}
        P3 & Australia & 51-60 & Female &  Senior Project Manager & Between 15-20 years & Healthcare, Professional Services, Government\\
        
        P4 & India & 20 - 30  & Male & Digital Accessibility Engineer, Software Developer  & Between 10-15 years & Healthcare, Insurance, Finance\\
        
        \rowcolor{tableGray}
        P5 & USA & 31-40 & Female & Senior QA Engineer & Between 5-10 years & Insurance\\
        
        P6 & Australia & 31-40 & Female & Senior Software Engineer & Between 10-15 years & Insurance, Finance\\

         \bottomrule
    \end{tabular}}
    
\end{table*}

\section{Final Version of the SE Empathy Scales} \label{sec:Final Version of the SE-Oriented Empathy Scales}

\subsection{Balancing Positively \& Negatively Worded Items} \label{sec:Balancing Positively-Worded and Negatively-Worded Items}
To improve the psychometric integrity of the scales and reduce response bias, we balanced the positively and negatively worded items across both scales. Positively worded items affirmed empathic behaviours or attitudes (e.g., ``I try to understand the actions and emotions of co-workers''), while negatively worded items indicated a lack of empathy (e.g., ``When working in a team, I rarely consider how things look from others' perspectives''). This approach aimed to minimise response patterns driven by bias rather than genuine reflection.
Negatively worded items are commonly used in personality and psychological measures to counter aberrant response behaviours and minimise the confounding effects of such responses \cite{paulhus1991measurement, weijters2013reversed}. They help address three common types of invalid responses: (1) acquiescence response style, a tendency to consistently agree or disagree with items regardless of their content; (2) careless responding, refers to random or inattentive answering without considering item meaning; and (3) confirmation bias, a tendency to selectively interpret statements in ways that affirm existing self-perceptions \cite{davies2003confirmatory}. For instance, when an item references extraversion, participants may recall situations in which they were outgoing; conversely, when the item references introversion, they may recall the opposite \cite{weijters2013reversed}.
As a result, the final scales included a curated mix of positively and negatively framed items, promoting thoughtful engagement and more accurate responses. The practitioner scale included 15 negatively worded and 22 positively worded items, while the user empathy scale contained 17 negatively worded and 20 positively worded items.

\subsection{Simplifying Double-Barrelled Items} \label{sec:Simplifying Double-Barrelled Items}
As part of the item refinement process for the final versions of the scales, we systematically reviewed and simplified double-barrelled items. These are the items that combine two or more ideas within a single statement. Such wording introduces ambiguity, as respondents may agree with one part of the item but disagree with the other, leading to unreliable or uninterpretable responses. For example, an item such as ``I do not adapt my approach based on users' preferences, needs, or feedback'' may confuse respondents who are comfortable acting on feedback but less confident addressing preferences directly.
To address this, we carefully deconstructed such items into separate, single-focus statements. This improved item clarity, reduced cognitive load, and increased the precision of measurement. Feedback from both experts and practitioners during the expert evaluation and cognitive interview phases consistently reinforced the need to simplify such items. Items flagged as unclear or conflated were reworded or split to ensure that each item measured only one distinct construct.
\textcolor{black}{Although double-barrelled items were identified and refined during earlier stages (Section \ref{sec:Expert Evaluation} \& \ref{sec:Cognitive Interviews}), scale development is inherently iterative, with different evaluation phases revealing different types of ambiguity. Such iterative refinement is common in established scale development studies, including the Jefferson Scale of Empathy \cite{hojat2001jefferson, hojat2016jefferson} and the EMPA-D scale \cite{drouet2024development, cabrera2010author}. Hence, we conducted a final targeted review to ensure that no double-barrelled items remained in the final instrument.} 
This approach improved both the content validity and interpretability of the scale, ensuring participant responses accurately reflected their empathic dispositions and behaviours in SE.

\subsection{Scoring of the SE Empathy Scales} \label{sec:Scoring of the SE Empathy Scales}
We employed a 7-point Likert scale to assess responses across the SE Empathy Scales. The scale ranged from 1 to 7, where 1 represented `Does not describe me at all' and 7 represented `Completely describes me'. The full set of scale points included: Does not describe me at all, Barely describes me, Somewhat describes me, Moderately describes me, Generally describes me, Mostly describes me, and Completely describes me.
Positively worded items, which indicated the presence of empathy, were scored in a standard manner (e.g., a rating of 7 denoted high agreement with empathic behaviour). Conversely, negatively worded items, which reflected a lack of empathy, were reverse-scored to maintain consistency in interpretation (e.g., a rating of 7 indicated low empathy when the respondent strongly agreed with a negatively phrased statement).
We opted for a Likert scale rather than a dichotomous response format (e.g., Yes/No, Agree/Disagree) because such scales allow for a greater range of item responses, thereby offering more nuanced variation and enhanced discriminatory power \cite{oppenheim2000questionnaire}. Further, likert scales often produce results that resembles a normal distribution \cite{likert1932technique}, enabling the use of parametric statistical techniques that assume evenly spaced responses and normally distributed data. In addition, we selected a 7-point scale instead of the more common 5-point scale to minimise the potential for response bias, such as a tendency to rely on extreme options. Prior research has shown that the two additional points on a 7-point scale may help reduce such biases and improve response precision \cite{polgar2011introduction, reynolds2017measurement}.

\subsection{Final Versions of the Scales} \label{sec:Final Versions of the Scales}
Following all prior stages of item development and refinement, we finalised the structure and wording of both empathy scales. To ensure the conceptual and linguistic precision of the refined items, we conducted two additional consultations with one of our original empathy experts (E5), primarily via email. These consultations focused on reviewing the most recent item revisions for conceptual clarity, redundancy, and overall coherence with the theoretical framework underpinning the scale. The expert provided feedback on item framing, the simplification of double-barrelled items, and language clarity. Based on these insights, we made several minor refinements, including rephrasing items, simplifying terminology, and resolving any remaining conceptual overlaps. The final scales consisted of two instruments: one measuring empathy among practitioners (EmpathiSEr-P) and the other assessing practitioner empathy towards users (EmpathiSEr-U). Each scale captured three dimensions of empathy, namely cognitive empathy, affective empathy, and empathic responses, and included a balanced mix of positively and negatively worded items. Both scales consisted of 37 items and formed the basis for the subsequent phase of psychometric validation. Statistical analysis was conducted using SPSS (version 29). 

\section{Psychometric Analyses of the Scales} \label{sec:Psychometric Analyses of the SE-Oriented Empathy Scales}
Psychometric analysis refers to the process of evaluating how well a scale measures what it is intended to measure \cite{cabrera2010author}. In this study, we applied psychometric analyses to assess the quality and reliability of the EmpathiSEr-P and EmpathiSEr-U scales. This included examining the scales' internal structure (e.g., factor analysis), reliability (e.g., internal consistency), and validity (e.g., how well the scales relate to established empathy measures). These analyses helps ensure that the scales are both scientifically sound and practically useful for assessing empathy in SE contexts.
To assess the psychometric properties of the final scales, we conducted a survey\footnote{Approved by Monash Human Research Ethics Committee. ERM Reference Number: 47675} with 229 software practitioners recruited via Prolific\footnote{https://www.prolific.com/} applying the same checks mentioned in Section \ref{sec:Initial Version and Pretest}. 
\textcolor{black}{Participants received monetary compensation at a rate of approximately 7.60 GBP per hour, consistent with the platform's recommended payment guidelines at the time of data collection.}
Table \ref{tab:Demographic Information of the Validation Survey Practitioners} provides a summary of the demographic information of these participants. The survey included demographic questions, final versions of the EmpathiSEr scales, and other validated scales (Table \ref{tab:Details of the Instruments used for Convergent and Discriminant Validity}). 

\begin{table*}[htbp]
    \caption{Overview of the Demographics of the Validation Survey Participants}
    \label{tab:Demographic Information of the Validation Survey Practitioners}
    \resizebox{\textwidth}{!}{
    \begin{tabular}{@{}llllll@{}}
        \toprule
        \textbf{Role} & \textbf{\# of Practitioners} & \textbf{Age} & \textbf{\# of Practitioners} &  \textbf{Experience} & \textbf{\# of Practitioners} \\
        \midrule
        
        Developer & \blackbar{59}{118} & 20-30 & \blackbar{30.5}{61} & 3-5 years & \blackbar{28}{56}\\
        
        Stakeholder & \blackbar{55.5}{111} & 31-40 & \blackbar{36.5}{73} & 5-10 years & \blackbar{39.5}{79} \\

        & & 41-50 & \blackbar{32}{64} & 10-15 years & \blackbar{17.5}{35}\\
        
        \cmidrule(r){1-2} 
        
        \textbf{Gender} & \textbf{\# of Practitioners} & 51-60 & \blackbar{12.5}{25} & 15-20 years & \blackbar{10.5}{21}\\

        \cmidrule(r){1-2} 
        
        Female & \blackbar{28.5}{57} & 61-70 & \blackbar{2}{4} & 20-30 years & \blackbar{13.5}{27}\\
        Male & \blackbar{84}{168} & Above 70 & \blackbar{1}{2} & 30-40 years & \blackbar{4.5}{9} \\
        Non Binary & \blackbar{1.5}{3} &  &  & 40-50 years & \blackbar{1}{2}\\
        Prefer not to say & \blackbar{0.5}{1} &  &  &  & \\
        
        \cmidrule(r){1-6} 

        \textbf{Team Size} & \textbf{\# of Practitioners} & \textbf{Org. Size} & \textbf{\# of Practitioners} & \textbf{Development Method Used} & \textbf{\# of Practitioners}\\
        
        \cmidrule(r){1-6} 

         No members & \blackbar{1.5}{3} & Self-Employed & \blackbar{3.5}{7} & Traditional (Waterfall) & \blackbar{75.5}{151}\\

         Less than or equal to 5 & \blackbar{24}{48} & Startup & \blackbar{3}{6} & Agile - Scrum & \blackbar{79.5}{159}\\

         5 - 10 & \blackbar{45.5}{91} & Small & \blackbar{16.5}{33} & Agile - Kanban & \blackbar{39}{78}\\

         10 -20 & \blackbar{30.5}{61} & Medium & \blackbar{26}{52} & XP & \blackbar{16.5}{33}\\

         More than 20 & \blackbar{13}{26} & Large & \blackbar{65.5}{131} &  & \\

         \cmidrule(r){1-6} 
         \textbf{Country} & \textbf{\# of Practitioners} & \textbf{Country} & \textbf{\# of Practitioners} & \textbf{Country} & \textbf{\# of Practitioners}  \\
         \cmidrule(r){1-6} 

         USA & \blackbar{34}{68} & France & \blackbar{1.5}{3} & Brazil & \blackbar{0.5}{1} \\
         UK & \blackbar{32.5}{65} & Germany & \blackbar{1.5}{3} & Chile & \blackbar{0.5}{1}  \\
         Portugal & \blackbar{8.5}{17} & Hungary & \blackbar{1.5}{3} & Finland & \blackbar{0.5}{1}  \\
         Canada & \blackbar{5.5}{11} & Netherlands & \blackbar{1.5}{3} & Ireland & \blackbar{0.5}{1}  \\
         Italy & \blackbar{5.5}{11} & Croatia & \blackbar{1}{2} & Israel & \blackbar{0.5}{1}  \\
         Poland  & \blackbar{5}{10} & Slovakia & \blackbar{1}{2} & Japan & \blackbar{0.5}{1}  \\
         Greece & \blackbar{3}{6} & Sweden & \blackbar{1}{2} & Latvia & \blackbar{0.5}{1} \\
         Mexico & \blackbar{2.5}{5} & Austria & \blackbar{0.5}{1} & New Zealand & \blackbar{0.5}{1} \\
         Spain & \blackbar{2}{4} & Belgium & \blackbar{0.5}{1} & Slovenia & \blackbar{0.5}{1} \\
         Australia & \blackbar{1.5}{3} &  & &  &  \\
            
        \bottomrule
    \end{tabular}
    }

\end{table*}

\subsection{Exploratory Factor Analysis} \label{sec:Exploratory Factor Analysis}
Exploratory factor analysis (EFA) is a statistical method used to uncover the underlying structure of a set of observed variables \cite{gorsuch1974factor}. It identifies latent constructs by grouping together items that are highly correlated, thereby suggesting they reflect the same underlying factor. EFA can also support instrument refinement by helping eliminate items with low or ambiguous loadings, retaining only those with sufficiently strong associations with a single factor, typically defined as factor loadings greater than $|0.30|$ \cite{gorsuch1974factor}. 

In this study, we conducted EFA to explore the latent structure of the EmpathiSEr scales and evaluate the extent to which the empirically derived factor groupings aligned with the theoretical constructs. The EFA examined solutions ranging from two to four factors using the principal axis factoring. Sampling adequacy was evaluated using the Kaiser-Meyer-Olkin (KMO) measure, where values greater than or equal to 0.6 are considered acceptable \cite{kaiser1974index}. The Bartlett Test of Sphericity was also used to ensure sufficient inter-item correlations for factor extraction \cite{tabachnick1996multivariate}. Given the potential for correlations among underlying factors, an oblique rotation method (direct oblimin) was employed \cite{tabachnick1996multivariate}.
The number of factors in the final EFA solution was determined by evaluating whether items loaded primarily onto a single factor (i.e., a simple structure) and by critically assessing the interpretability and theoretical coherence of each factor. Factors containing conceptually unrelated items were treated as evidence of poor construct validity, prompting further refinement or re-evaluation of the model, regardless of statistical indicators.

\subsubsection{EFA Results for the EmpathiSEr-P Scale} \label{sec:EFA Results for the EmpathiSEr-P Scale}
We conducted an EFA using principal axis factoring on the 37 items of the EmpathiSEr-P scale, applying direct oblimin rotation to allow for correlations between factors. Prior to analysis, the suitability of the data for EFA was assessed. The sample size (n = 229) provided a case-to-item ratio of approximately 6:1, satisfying the commonly recommended minimum of 5:1 or at least 200 cases \cite{howard2016review}. The overall Kaiser-Meyer-Olkin (KMO) measure of sampling adequacy was 0.915, which is classified as ``marvellous'' according to Kaiser's criterion \cite{kaiser1974index}, indicating excellent suitability of the data for factor analysis. Bartlett's Test of Sphericity was also statistically significant, X$^2$(666) = 4423.587, p $<0.001$, indicating that the data were likely factorisable.

The EFA initially identified seven factors with eigenvalues greater than 1. However, visual inspection of the scree plot suggested a four-factor solution, which accounted for 52\% of the variance. This indicated that two-, three-, and four-factor solutions were all plausible. We conducted EFA iteratively across these three solutions to identify a simple structure, i.e., a solution where each item loaded significantly on only one factor. We applied a combination of criteria to remove cross-loading items based on recommended cut-offs \cite{howard2016review}. Specifically, items with cross-loadings of $|0.30|$ or above on multiple factors were removed.
Once a simple structure was achieved, we examined the pattern of item loadings within each factor to interpret their conceptual coherence, guided by empirical relationships among items. Items that did not align meaningfully with the emerging factor themes were removed, and EFA was re-run to confirm the absence of cross-loadings in the revised solution.

For the EmpathiSEr-P scale, the data did not support both four- and two-factor solutions. While they yielded simple structures, the factor interpretations were conceptually incoherent. In contrast, the three-factor solution offered both a simple structure and interpretable factors. This solution, therefore, was selected as the final EFA model.
The final structure comprised 20 items loading onto three components: cognitive empathy (CE), affective empathy (AE), and empathic responses (ER). The Kaiser-Meyer-Olkin (KMO) measure of sampling adequacy was 0.860, which is considered ``meritorious'' \cite{kaiser1974index}, and Bartlett's Test of Sphericity was statistically significant, X$^2$(190) = 1647.348, p $<0.001$. The three-factor solution explained 49\% of the total variance, with factor loadings ranging from 0.316 to 0.762. The pattern matrix with final factor loadings is presented in Table \ref{tab:The Final Pattern Matrix of EmpathiSEr-P Scale}. Additional details related to the final three-factor solution, details regarding the EFA of the initial three-factor solution with all 37 items, and scoring key of Empathiser-P scale are provided in our online Appendix\footnote{https://github.com/Hashini-G/SupplementaryInfoPackage-EmpathiSErScales}.

Estimates of internal consistency for the EmpathiSEr-P total scale and its three factors were assessed using Cronbach's alpha (Table \ref{tab:Pearson Correlations between EmpathiSEr-P Final Scale with Other Validity Instruments}). Alpha coefficients of 0.851, 0.887, 0.680, and 0.736 were obtained for the total scale, cognitive empathy, affective empathy, and empathic responses, respectively. These values indicate good to acceptable levels of internal consistency. Generally, Cronbach's alpha values above 0.70 are considered adequate \cite{cortina1993coefficient}. The alpha of 0.680 for the affective empathy factor is considered acceptable given that it comprises only three items, and reliability coefficients tend to be lower with fewer items \cite{cortina1993coefficient}.


\begin{table*} [htbp]
    \scriptsize
    \centering
    \caption{Pattern Matrix for the Final EmpathiSEr-P Scale (Direct Oblimin Rotation)\\ The pattern matrix shows how strongly each item relates to a specific empathy factor (cognitive, affective, or behavioural), while controlling for the influence of other factors. It helps identify which items best represent each dimension of empathy.}
    \label{tab:The Final Pattern Matrix of EmpathiSEr-P Scale}
     \resizebox{\textwidth}{!}{
    \begin{tabular}{P{0.07\textwidth} P{0.78\textwidth} L{0.05\textwidth} P{0.05\textwidth} P{0.05\textwidth}}
    \toprule
         \textbf{ID} & \textbf{Item} & \multicolumn{3}{c}{\textbf{Factors}} \\
         \cmidrule(r){3-5} 
          &  & \centering \textbf{1} &  \textbf{2} &  \textbf{3} \\
         \midrule
         
         emp\textunderscore p1 & I do not actively consider co-workers' perspectives or work-related needs when developing software solutions. & \textminus0.469  &  & \\
         
         emp\textunderscore p2 & I make an effort to understand the work-related needs and concerns of co-workers, by considering how things look from their perspective. & 0.650 &  & \\   
         
         emp\textunderscore p7 & I pay attention to co-workers' work styles \& communication preferences to better understand their perspective. & 0.708 &  & \\
         
         emp\textunderscore p8 & I understand the technical and domain-specific challenges co-workers face by considering their context, including their individual strengths and weaknesses. & 0.668 &  & \\
         
         emp\textunderscore p11 & I take co-workers' perspectives into account when discussing different viewpoints during meetings, such as code reviews or design sessions. & 0.762 &  & \\
         
         emp\textunderscore p12 & I try to understand the concerns of co-workers about proposed solutions by taking their perspective into account. & 0.724 &  & \\
         
         emp\textunderscore p20 & I try to understand the actions and emotions of co-workers. & 0.685 &  & \\
         
         emp\textunderscore p22 & I clarify co-workers' doubts about requirements or domain knowledge by taking their perspective into account. & 0.719 &  & \\
         
         emp\textunderscore p23 & I try to understand ethnic \& cultural differences that influence the communication and behaviour of co-workers. & 0.648 &  & \\

         emp\textunderscore p13 & During my interactions with co-workers, I am emotionally affected due to their emotional tension, stress, and frustration. & & \textminus0.604 & \\ 
         
         emp\textunderscore p15 & I feel uneasy during conflicts, disagreements, or tension among co-workers. & & \textminus0.557 & \\
         
         emp\textunderscore p16 & I am emotionally sensitive to the emotional states of co-workers during high-pressure project situations, such as tight deadlines or significant challenges. & & \textminus0.704 &  \\

         emp\textunderscore p5 & I find it difficult to remain flexible when co-workers face personal emergencies, especially if it affects scheduled tasks. & &  & 0.318  \\
         
         emp\textunderscore p9 & When co-workers point out issues in my work, I usually focus on defending my choices rather than understanding their perspective. & &  & 0.377 \\
         
         emp\textunderscore p10 & I tend to give feedback without considering how it might be perceived from my co-workers' perspective. & &  & 0.743 \\
         
         emp\textunderscore p14 & I tend to provide feedback without adjusting my tone or approach based on how it might affect co-workers. & &  & 0.672 \\
         
         emp\textunderscore p21 & I don't adjust my communication style based on co-workers' roles or level of technical expertise. & &  & 0.618 \\
         
         emp\textunderscore p25 & I tend to make decisions without fully considering all relevant perspectives or all sides of a disagreement. & &  & 0.493 \\
         
         emp\textunderscore p28 & When working in a team, I don't always make an effort to ensure everyone feels heard and understood. & &  & 0.316 \\
         
         emp\textunderscore p34 & When assigning work, I do not consider co-workers' expertise, capacity, or capability. & &  & 0.346 \\
   
         \bottomrule
    \end{tabular}}
    \begin{flushleft}
        \textit{Note: Factor loadings $<0.30$ are constrained. Factor 1: Cognitive Empathy; Factor 2: Affective Empathy; Factor 3: Empathic Responses.} 
    \end{flushleft}
\end{table*}

\subsubsection{EFA Results for the EmpathiSEr-U Scale} \label{sec:EFA Results for the EmpathiSEr-U Scale}
An EFA was also conducted on the 37 items of the EmpathiSEr-U scale using principal axis factoring with direct oblimin rotation. Preliminary checks confirmed data suitability for EFA: the sample size (n = 229) met recommended thresholds \cite{howard2016review}, the KMO value (0.946) was classified as ``marvellous'' \cite{kaiser1974index} indicating excellent suitability for EFA, and Bartlett's Test of Sphericity was significant, X$^2$(666) = 5268.113, p $<0.001$.

Initial extraction yielded five factors with eigenvalues greater than 1, but the scree plot indicated a possible four-factor solution. As with EmpathiSEr-P, we explored two-, three-, and four-factor solutions to identify the most parsimonious and interpretable model. Items with significant cross-loadings ($>=|0.30|$) or weak conceptual fit were iteratively removed. The four- and two-factor models failed to yield coherent factor structures.
The three-factor solution offered the best balance of simple structure and conceptual clarity. This final model consisted of 18 items across three factors: cognitive empathy (CE), affective empathy (AE), and empathic responses (ER), explaining 58\% of the total variance. Factor loadings ranged from 0.393 to 0.846. The final KMO remained high (0.913), and Bartlett's Test was again significant, X$^2$(171) = 2128.475, p $<0.001$. However, during the EFA, one item (emp\textunderscore u20) showed minor cross-loading (0.395 on ER, 0.300 on CE). It was retained due to its stronger loading on the ER factor and its clear conceptual fit with that construct. Removing it would have reduced the number of ER items and weakened its internal consistency. With this item included, the ER factor consisted of four items and demonstrated acceptable internal consistency for exploratory research. Nonetheless, future studies should further assess this item in larger and more diverse samples to assess the need for refinement or replacement.
The final pattern matrix is presented in Table \ref{tab:The Final Pattern Matrix of EmpathiSEr-U Scale}. Additional details related to the final three-factor solution, details regarding the EFA of the initial three-factor solution with all 37 items, and scoring key of Empathiser-U scale are provided in our online Appendix\footnotemark[\value{footnote}].

Internal consistency estimates were strong: Cronbach's alpha values were 0.899 for the total scale, 0.887 for CE, 0.896 for AE, and 0.627 for ER (Table \ref{tab:Pearson Correlations between EmpathiSEr-U Final Scale with Other Validity Instruments}). While the ER factor's alpha is slightly below the conventional threshold of 0.70, it is acceptable given the small number of items \cite{cortina1993coefficient}.


\begin{table*} [htbp]
    \scriptsize
    \centering
    \caption{Pattern Matrix for the Final EmpathiSEr-U Scale (Direct Oblimin Rotation)}

    \label{tab:The Final Pattern Matrix of EmpathiSEr-U Scale}
     \resizebox{\textwidth}{!}{
    \begin{tabular}{P{0.07\textwidth} P{0.78\textwidth} L{0.05\textwidth} P{0.05\textwidth} P{0.05\textwidth}}
    \toprule
         \textbf{ID} & \textbf{Item} & \multicolumn{3}{c}{\textbf{Factors}} \\
         \cmidrule(r){3-5} 
          &  & \centering \textbf{1} & \textbf{2} & \textbf{3} \\
         \midrule
         
         emp\textunderscore u2 & I make an effort to understand the needs and concerns of users, by considering how things look from their perspective. & 0.808  &  & \\
         
         emp\textunderscore u3 & I consider users' needs to guide development decisions, rather than focusing solely on completing tasks. & 0.779 &  & \\   
         
         emp\textunderscore u5 & I do not make an effort to understand the technical limitations and other challenges users face from their perspective. & \textminus0.461 &  & \\
         
         emp\textunderscore u7 & I try to understand the concerns of users about our solutions by considering their perspective and feedback. & 0.758 &  & \\
         
         emp\textunderscore u17 & I don't usually consider the users' perspective to understand their experience of using the software. & \textminus0.639 &  & \\
                  
         emp\textunderscore u23 & I clarify users' doubts about the software by taking their perspective into account. & 0.706 &  & \\
         
         emp\textunderscore u24 & I consider users' needs and perspective when creating documentation to ensure it is easy to understand. & 0.757 &  & \\
         
         emp\textunderscore u33 & When working with users, I don't always try to understand how things look from their perspective. & \textminus0.577 &  & \\

         emp\textunderscore u9 & I get emotionally affected when users express negative emotions due to issues in our software. & & 0.846 & \\ 
         
         emp\textunderscore u11 & I often become emotionally engaged with users' feelings during conflicts or disagreements. & & 0.767 & \\
         
         emp\textunderscore u14 & I am emotionally affected when users face challenges while using the software we developed. & & 0.806 &  \\

         emp\textunderscore u16 & I am emotionally moved when users share personal stories about how our software has helped them. & & 0.486 &   \\
         
         emp\textunderscore u19 & I get emotionally involved with the users' experiences and emotional responses of using our software. & & 0.654 &  \\
         
         emp\textunderscore u26 & I get deeply involved with the feelings of users during failures in our software. & & 0.776 &  \\
         
         emp\textunderscore u10 & I don't usually take users' experiences or emotional states into account when responding to them. & &  & 0.393 \\
         
         emp\textunderscore u20 & I don't make an effort to build a human-level connection with users. & &  & 0.395 \\
         
         emp\textunderscore u25 & Generally I am able to make decisions based on my knowledge and without being overly influenced by emotions of users. & &  & 0.433 \\
         
         emp\textunderscore u27 & I rarely reflect on my interactions with users to understand their emotions and perspectives. & &  & 0.472 \\
            
         \bottomrule
    \end{tabular}}
    \begin{flushleft}
        \textit{Note: Factor loadings $<0.30$ are constrained. Factor 1: Cognitive Empathy; Factor 2: Affective Empathy; Factor 3: Empathic Responses.}
    \end{flushleft}
\end{table*}

\subsection{Convergent and Discriminant Validity} \label{sec:Convergent and Discriminant Validity of EmpathiSEr Scales}
Additional scales were used to assess the \textit{convergent} and \textit{discriminant} validity of the EmpathiSEr scales. Evidence supporting its validity includes significant positive correlations with measures conceptually related to empathy, no significant correlations with measures unrelated to empathy, and significant negative correlations with constructs known to hinder empathic engagement. To minimise bias in responses, these validation scales were presented after the EmpathiSEr items, and participants were not able to return to previous responses once answered.
\textbf{To assess convergent validity}, we administered the four subscales of the IRI (Perspective Taking, Empathic Concern, Personal Distress, and Fantasy) \cite{davis1980multidimensional}, the EQ \cite{baron2004empathy}, and the three subscales of the EMPA-D scale (Emotional interest and Perspective taking, Personal experience, and Self-awareness) \cite{drouet2024development}. We also included several single-item measures of personal attributes, specifically empathy, compassion, and sense of humour. Each attribute was clearly defined in the instrument. Respondents rated the extent to which they currently possessed each attribute on a 100-point scale. We predicted significant low to moderate positive correlations between the EmpathiSEr scales and these instruments and attributes.
\textbf{For discriminant validity}, existing literature has shown that narcissism is negatively associated with empathy \cite{hepper2014moving, burgmer2021don}. Based on this, we administered Single Item Narcissism Scale (SINS) \cite{konrath2014development} and Narcissistic Personality Inventory (NPI-13) \cite{gentile2013test}. We also included Balanced Inventory of Desirable Responding (BIDR-16) \cite{hart2015balanced} to address potential response bias. We predicted negative correlations between the EmpathiSEr scales and these measures.
The details of all additional instruments and attributes are outlined in Table \ref{tab:Details of the Instruments used for Convergent and Discriminant Validity}.

\begin{table*}[htbp]
    \scriptsize
    \centering
    \setlength{\aboverulesep}{0pt}
    \setlength{\belowrulesep}{0pt}
    \setlength{\extrarowheight}{.5ex}
    \caption{Details of the Instruments used for Convergent and Discriminant Validity}
    \label{tab:Details of the Instruments used for Convergent and Discriminant Validity}
     \resizebox{\textwidth}{!}{
    \begin{tabular}{P{0.2\textwidth} P {0.8\textwidth}}
    \toprule
     \textbf{Instrument/Attribute}  &  \textbf{Description}\\
     \midrule
      \rowcolor{tableGray}
      Interpersonal Reactivity Index (IRI) \cite{davis1980multidimensional, davis1983measuring}   & All four subscales of the IRI were used (Cronbach's $\alpha$ = 0.883): Perspective Taking (PT), which measures the tendency to adopt others' viewpoints; Empathic Concern (EC), which captures feelings of sympathy and concern for others; Personal Distress (PD), reflecting self-oriented feelings of anxiety in tense interpersonal situations; and the Fantasy Scale (FS), which assesses the tendency to imaginatively identify with fictional characters. Each subscale contains seven items rated on a 5-point Likert scale ranging from 1 - ``Does not describe me well'' to 5 - ``Describes me very well.'' Factor scores for the total scale and each subscale were computed by calculating the mean, in accordance with the original scoring guidelines.\\
         
      Empathy Quotient (EQ) \cite{baron2004empathy}   & The EQ is a self-administered questionnaire designed to measure empathy in adults. Respondents rated their agreement with 40 statements using a 4-point Likert scale ranging from 1 – ``Strongly agree'' to 4 – ``Strongly disagree.'' Following reliability analysis (Cronbach's $\alpha$ = 0.909), a factor score was computed for the total EQ scale.\\

       \rowcolor{tableGray}
       Empathy in Design Scale (EMPA-D) \cite{drouet2024development}  & All subscales of the EMPA-D were used (Cronbach's $\alpha$ = 0.933). The subscales include Emotional Interest and Perspective Taking (EIPT), which assesses employees' willingness to learn from users, including their curiosity and interest in users; Personal Experience (PE), which evaluates the ability to draw on personal or acquaintances' experiences to understand users' perspectives; and Self-Awareness (SA), which captures the ability to differentiate between one's own experience and that of users. EIPT comprises six items, PE includes two items, and SA consists of three items. All items are rated on a 7-point Likert scale ranging from 1 - ``Does not describe me at all'' to 7 - ``Completely describes me.'' Factor scores for the total scale and each subscale were calculated by taking the mean, following the original scoring guidelines.\\

        Single Item Narcissism Scale (SINS) \cite{konrath2014development} & The SINS consists of a single item where participants rate the statement ``I am a narcissist'' on a 7-point Likert scale ranging from 1 – ``Not very true of me'' to 7 – ``Very true of me.'' A clarification is provided to participants that ``narcissist'' refers to being egotistical, self-focused, and vain. \\

        \rowcolor{tableGray}
         Narcissistic Personality Inventory (NPI-13) \cite{gentile2013test} & This widely used measure of trait narcissism includes three subscales: Leadership/Authority (LA), Grandiose/Exhibitionism (GE), and Entitlement/Exploitativeness (EE). We used the 13-item short form to reduce participant fatigue, with all subscales included (Cronbach's $\alpha$ = 0.768). For each item, respondents selected the statement they agreed with most. A factor score was computed for the overall scale following reliability analysis. Although the EE subscale has shown relatively low internal consistency in prior research \cite{ackerman2011does}, it remains a meaningful divergent measure for empathy. \\

         Balanced Inventory of Desirable Responding (BIDR-16) \cite{paulhus1991measurement, hart2015balanced} & The BIDR-16 was used to account for response bias and due to its theoretical relevance to empathy. It includes two subscales: Self-Deceptive Enhancement (SDE), which reflects honest but overly positive self-reporting, and Impression Management (IM), which reflects deliberate efforts to present oneself favourably. Participants rated each item on a 7-point Likert scale ranging from 1 – ``Not very true'' to 7 – ``Very true.'' Following reliability analysis (Cronbach's $\alpha$ = 0.833), a factor score was computed. \\

         \rowcolor{tableGray}
         Empathy & Defined as ``the ability to understand and relate to the perspectives, challenges, and emotions of users, teammates, or stakeholders.'' \\

          Compassion & Defined as ``taking action to support or help others in response to their challenges or difficulties.'' \\

         \rowcolor{tableGray}
          Sense of Humour & Defined as ``the ability to use or appreciate light-heartedness and appropriate humour to build rapport, ease stress, and enhance collaboration within technical teams.''\\
         
    \bottomrule
    \end{tabular}}
\end{table*}

\subsubsection{Convergent \& Discriminant Validity of EmpathiSEr-P} \label{sec:Convergent and Discriminant Validity Results for the EmpathiSEr-P Scale}

The correlations between EmpathiSEr-P scale and other external instruments are displayed in Table \ref{tab:Pearson Correlations between EmpathiSEr-P Final Scale with Other Validity Instruments}.
We found significant positive associations between the EmpathiSEr-P Total score and validated empathy measures, including EQ (r = 0.633, p $<0.001$), IRI (r = 0.646, p $<0.001$), and EMPA-D (r = 0.610, p $<0.001$) supporting the convergent validity of EmpathiSEr-P scale as a distinct empathy measure. IRI subscales were also positively correlated with EmpathiSEr-P subscales, with the strongest correlation emerged between IRI PT and EmpathiSEr-P CE (r = 0.661, p $<0.001$). As expected, EMPA-D subscales were also positively associated with EmpathiSEr-P subscales, with strong links between EMPA-D EIPT and EmpathiSEr-P CE (r = 0.586, p $<0.001$), and EMPA-D PE and EmpathiSEr-P CE (r = 0.621, p $<0.001$). We also observed  positive correlations between EmpathiSEr-P Total and the personal attributes including empathy (r = 0.564, p $<0.001$), compassion (r = 0.550, p $<0.001$) and humour (r = 0.192, p = 0.004).

To evaluate discriminant validity, we predicted significant negative correlations between EmpathiSEr-P Total and established measures of narcissism (SINS and the NPI-13 EE subscale). As expected, SINS showed small negative correlations with EmpathiSEr-P Total (r = \textminus0.146, P = 0.027) and its AE subscale (r = \textminus0.166, P = 0.012), but no significant associations with the CE or ER subscales. Similarly, NPI-13 EE subscale showed small but significant negative correlations with EmpathiSEr-P Total (r = \textminus0.284, P $<0.001$), CE (r = \textminus0.245, P $<0.001$), and AE (r = \textminus0.243, P $<0.001$). These results support the scale's discriminant validity, confirming that it captures constructs distinct from narcissism.

We examined the extent to which desirable responding influenced responses to the EmpathiSEr-P scale, using BIDR-16. We found negligible but statistically significant associations between EmpathiSEr-P Total and BIDR-16 Total (r = 0.264, P $<0.001$), and BIDR-16 IM (r = 0.327, P $<0.001$). Similar patterns were observed between BIDR-16 Total and EmpathiSEr-P subscales: CE (r = 0.347, P $<0.001$), AE (r = 0.221, P $<0.001$), and a small negative association with ER (r = \textminus0.137, P = 0.040). No significant associations were found between BIDR-16 SDE and either EmpathiSEr-P Total or AE. However, SDE showed a negligible positive correlation with CE (r = 0.261, P $<0.001$) and a small negative association with ER (r = \textminus0.322, P $<0.001$). These findings suggest that responses to the EmpathiSEr-P were not strongly influenced by socially desirable responding, supporting the scale's credibility as a measure of self-reported empathy.

\begin{table*} [htbp]
    \footnotesize
    \centering
    \caption{Pearson Correlations between EmpathiSEr-P Final Scale with Other Instruments}
    \label{tab:Pearson Correlations between EmpathiSEr-P Final Scale with Other Validity Instruments}
    \resizebox{\textwidth}{!}{
    \begin{tabular}{llllllllllllllllllllllllll}
        \toprule
          & \textbf{1} & \textbf{2} & \textbf{3} & \textbf{4} & \textbf{5} & \textbf{6} & \textbf{7} & \textbf{8} & \textbf{9} & \textbf{10} & \textbf{11} & \textbf{12} & \textbf{13} & \textbf{14} & \textbf{15} & \textbf{16} & \textbf{17} & \textbf{18} & \textbf{19} & \textbf{20} & \textbf{21} & \textbf{22} & \textbf{23} & \textbf{24} & \textbf{25}\\
         \midrule
         1 & \textbf{(0.851)} &  &  &  &  &  &  &  &   &  &  &  &  &  &  &  &  &  &  &  &  &  &  &  & \\
         2 & 0.865** & \textbf{(0.887)} &  &  &  &  &  &  &   &  &  &  &  &  &  &  &  &  &  &  &  &  &  &  & \\
         3 & 0.756** & 0.428** & \textbf{(0.680)} &  &  &  &  &  &   &  &  &  &  &  &  &  &  &  &  &  &  &  &  &  & \\
         4 & 0.533** & 0.336** & 0.133* & \textbf{(0.736)} &  &  &  &  &   &  &  &  &  &  &  &  &  &  &  &  &  &  &  &  & \\
         5 & 0.633** & 0.619** & 0.461** & 0.223** & \textbf{(0.909)} &  &  &  &   &  &  &  &  &  &  &  &  &  &  &  &  &  &  &  & \\
         6 & 0.646** & 0.540** & 0.408** & 0.519** & 0.569** & \textbf{(0.883)} &  &  &   &  &  &  &  &  &  &  &  &  &  &  &  &  &  &  & \\
         7 & 0.695** & 0.661** & 0.512** & 0.270** & 0.700** & 0.689** & \textbf{(0.805)} &  &   &  &  &  &  &  &  &  &  &  &  &  &  &  &  &  & \\
         8 & 0.664** & 0.575** & 0.469** & 0.407** & 0.670** & 0.841** & 0.677** & \textbf{(0.867)} &   &  &  &  &  &  &  &  &  &  &  &  &  &  &  &  & \\
         9 & 0.412** & 0.331** & 0.234** & 0.405** & 0.352** & 0.766** & 0.347** & 0.509** & \textbf{(0.779)}  &  &  &  &  &  &  &  &  &  &  &  &  &  &  &  & \\
         10 & 0.095 & 0.002 & \textminus0.029 & 0.384** & \textminus0.070 & 0.547** & 0.000 & 0.214** & 0.296**  & \textbf{(0.811)} &  &  &  &  &  &  &  &  &  &  &  &  &  &  & \\
         11 & 0.610** & 0.643** & 0.425** & 0.153* & 0.649** & 0.563** & 0.652** & 0.636** & 0.362**  & \textminus0.017 & \textbf{(0.933)} &  &  &  &  &  &  &  &  &  &  &  &  &  & \\
         12 & 0.540** & 0.586** & 0.378** & 0.098 & 0.594** & 0.520** & 0.607** & 0.594** & 0.338**  & \textminus0.032 & 0.954** & \textbf{(0.963)} &  &  &  &  &  &  &  &  &  &  &  &  & \\
         13 & 0.587** & 0.621** & 0.361** & 0.225** & 0.586** & 0.629** & 0.584** & 0.661** &  0.436** & 0.123 & 0.818** & 0.715** & \textbf{(0.847)} &  &  &  &  &  &  &  &  &  &  &  & \\
         14 & 0.433** & 0.413** & 0.341** & 0.129 & 0.449** & 0.266** & 0.426** & 0.328** & 0.130*  & \textminus0.100 & 0.671** & 0.474** & 0.402** & \textbf{(0.781)} &  &  &  &  &  &  &  &  &  &  & \\
         15 & \textminus0.128 & \textminus0.009 & \textminus0.156* & \textminus0.172** & 0.017 & \textminus0.042 & \textminus0.066 & \textminus0.080 & 0.075  & \textminus0.054 & 0.068 & 0.102 & 0.059 & \textminus0.050 & \textbf{(0.768)} &  &  &  &  &  &  &  &  &  & \\
         16 & \textminus0.284** & \textminus0.245** & \textminus0.243* & \textminus0.105 & \textminus0.259** & \textminus0.181** & \textminus0.299** & \textminus0.272** & \textminus0.024  & 0.061 & \textminus0.134* & \textminus0.113 & \textminus0.178** & \textminus0.064 & 0.611** & \textbf{(0.476)} &  &  &  &  &  &  &  &  & \\
         17 & \textminus0.068 & 0.040 & \textminus0.069 & \textminus0.208** & 0.095 & \textminus0.043 & 0.022 & \textminus0.017 & 0.037  & \textminus0.159* & 0.133* & 0.157* & 0.107 & 0.016 & 0.852** & 0.404** & \textbf{(0.620)} &  &  &  &  &  &  &  & \\
         18 & \textminus0.004 & 0.108 & \textminus0.086 & \textminus0.088 & 0.121 & 0.072 & 0.048 & 0.038 & 0.129  & \textminus0.011 & 0.107 & 0.139* & 0.140* & \textminus0.068 & 0.817** & 0.190** & 0.540** & \textbf{(0.703)} &  &  &  &  &  &  & \\
         
         19 & \textminus0.146* & \textminus0.084 & \textminus0.166* & \textminus0.067 & \textminus0.136* & \textminus0.031 & \textminus0.121 & \textminus0.113 & 0.107  & 0.025 & \textminus0.096 & \textminus0.091 & \textminus0.088 & \textminus0.056 & 0.233** & 0.235** & 0.148* & 0.154* & \textbf{(-)} &  &  &  &  &  & \\
        
         20 & 0.264** & 0.347** & 0.221** & \textminus0.137* & 0.392** & 0.040 & 0.329** & 0.187** & \textminus0.088  & \textminus0.276** & 0.325** & 0.292** & 0.313** & 0.219** & 0.036 & \textminus0.301** & 0.160* & 0.130* & \textminus0.130 & \textbf{(0.833)} &  &  &  &  & \\
         
         21 & 0.106 & 0.261** & 0.099 & \textminus0.322** & 0.334** & \textminus0.137* & 0.159* & 0.012 & \textminus0.155*  & \textminus0.372** & 0.273** & 0.275** & 0.236** & 0.137* & 0.234** & \textminus0.122 & 0.289** & 0.285* & \textminus0.069 & 0.821** & \textbf{(0.778)} &  &  &  & \\
         
         22 & 0.327** & 0.318** & 0.265** & 0.073 & 0.325** & \textminus0.188** & 0.383** & 0.288** &  \textminus0.001 & \textminus0.106 & 0.273** & 0.218** & 0.287** & 0.227** & \textminus0.154* & \textminus0.371** & \textminus0.006 & \textminus0.050 & \textminus0.145* & 0.856** & 0.408** & \textbf{(0.801)} &  &  & \\
         
         23 & 0.564** & 0.591** & 0.340** & 0.237** & 0.645** & 0.569** & 0.629** & 0.632** & 0.320** & 0.065 &  0.663**& 0.660** & 0.575** & 0.346** & 0.007 & \textminus0.214** & 0.044 & 0.116 & \textminus0.117 & 0.229** & 0.158* & 0.223** & \textbf{(-)} &  & \\
        
         24 & 0.550** & 0.519** & 0.372** & 0.279** & 0.570** & 0.558** & 0.594** & 0.683** & 0.279**  & 0.054 & 0.611** & 0.572** & 0.591** & 0.350** & \textminus0.047 & \textminus0.261** & 0.017 & 0.064 & \textminus0.139* & 0.249** & 0.100 & 0.308** & 0.704** & \textbf{(-)} & \\
         
         25 & 0.192** & 0.218** & 0.080 & 0.106 & 0.224** & 0.098 & 0.187** & 0.155** & 0.100  & \textminus0.148* & 0.220** & 0.174** & 0.187** & 0.231** & \textminus0.052 & \textminus0.104 & 0.011 & \textminus0.042 &  0.024 & 0.032 & 0.054 & 0.002 & 0.192** & 0.217** & \textbf{(-)} \\

         \bottomrule
    \end{tabular}}
    \begin{flushleft}
        \textit{**Correlation is significant at the 0.01 level (two-tailed), *Correlation is significant at the 0.05 level (two-tailed). Cronbach's alpha values are in the diagonal within the brackets. 1: EmpathiSEr-P Total, 2: EmpathiSEr-P CE, 3: EmpathiSEr-P AE, 4: EmpathiSEr-P ER, 5: EQ-40, 6: IRI Total, 7: IRI PT, 8: IRI EC, 9: IRI FS, 10: IRI PD, 11: EMPA-D Total, 12: EMPA-D EIPT, 13: EMPA-D PE, 14: EMPA-D SA, 15: NPI-13 Total, 16: NPI-13 EE, 17: NPI-13 LA, 18: NPI-13 GE, 19: SINS, 20: BIDR-16 Total, 21: BIDR-16 SDE, 22: BIDR-16 IM, 23: Empathy (Personal Attribute), 24: Compassion, 25: Humour.}
    \end{flushleft}
\end{table*}

\subsubsection{Convergent \& Discriminant Validity of EmpathiSEr-U} \label{sec:Convergent and Discriminant Validity Results for the EmpathiSEr-U Scale}

The correlations between EmpathiSEr-U scale and other external instruments are displayed in Table \ref{tab:Pearson Correlations between EmpathiSEr-U Final Scale with Other Validity Instruments}. The EmpathiSEr-U Total showed strong positive associations with validated empathy scales, including EQ (r = 0.611, p $<0.001$), IRI (r = 0.693, p $<0.001$), and EMPA-D (r = 0.757, p $<0.001$), supporting its convergent validity. IRI subscales also correlated with EmpathiSEr-U subscales, with particularly strong links between IRI PT and EmpathiSEr-U CE (r = 0.626, p $<0.001$) and IRI EC and EmpathiSEr-U AE (r = 0.618, p $<0.001$). Similarly, EMPA-D subscales showed strong alignment, especially between EMPA-D EIPT and EmpathiSEr-U CE (r = 0.749, p $<0.001$) and EMPA-D PE and EmpathiSEr-U CE (r = 0.633, p $<0.001$). In addition, we observed  a positive correlation between EmpathiSEr-P Total and the personal attributes including Empathy (r = 0.595, p $<0.001$), compassion (r = 0.582, p $<0.001$) and humour (r = 0.148, p = 0.025).

Discriminant validity was supported by the negative correlations with narcissism measures. As expected, SINS correlated negatively with the EmpathiSEr-U Total (r = \textminus0.109, p = 0.101) and its CE subscale (r = \textminus0.181, p = 0.006), but no significant associations with the AE or ER subscales. Similarly, NPI-13 EE showed significant but small negative associations with EmpathiSEr-U Total (r = \textminus0.198, p = 0.003) and CE (r = \textminus0.245, p $<0.001$). These findings suggest the scale captures empathy constructs that are distinct from narcissistic traits.

We assessed the influence of socially desirable responding on EmpathiSEr-U using BIDR-16. Small but significant correlations were found between EmpathiSEr-U Total and BIDR-16 Total (r = 0.310, P $<0.001$), BIDR-16 SDE (r = 0.209, P $<0.001$), and BIDR-16 IM (r = 0.306, P $<0.001$). Similar subscale-level associations were observed, with the strongest link between CE and BIDR-16 Total (r = 0.350, p $<0.001$). BIDR-16 SDE was only associated with CE (r = 0.294, p $<0.001$). 
These results suggest limited influence of desirable responding, supporting the credibility of the EmpathiSEr-U as a self-report measure.



\begin{table*} [htbp]
    \footnotesize
    \centering
    \caption{Pearson Correlations between EmpathiSEr-U Final Scale with Other Instruments}
    \label{tab:Pearson Correlations between EmpathiSEr-U Final Scale with Other Validity Instruments}
    \resizebox{\textwidth}{!}{
    \begin{tabular}{llllllllllllllllllllllllll}
        \toprule
          & \textbf{1} & \textbf{2} & \textbf{3} & \textbf{4} & \textbf{5} & \textbf{6} & \textbf{7} & \textbf{8} & \textbf{9} & \textbf{10} & \textbf{11} & \textbf{12} & \textbf{13} & \textbf{14} & \textbf{15} & \textbf{16} & \textbf{17} & \textbf{18} & \textbf{19} & \textbf{20} & \textbf{21} & \textbf{22} & \textbf{23} & \textbf{24} & \textbf{25}\\
         \midrule
         1 & \textbf{(0.899)} &  &  &  &  &  &  &  &   &  &  &  &  &  &  &  &  &  &  &  &  &  &  &  & \\
         2 & 0.805** & \textbf{(0.887)} &  &  &  &  &  &  &   &  &  &  &  &  &  &  &  &  &  &  &  &  &  &  & \\
         3 & 0.841** & 0.420** & \textbf{(0.896)} &  &  &  &  &  &   &  &  &  &  &  &  &  &  &  &  &  &  &  &  &  & \\
         4 & 0.763** & 0.519** & 0.509** & \textbf{(0.627)} &  &  &  &  &   &  &  &  &  &  &  &  &  &  &  &  &  &  &  &  & \\
         5 & 0.611** & 0.604** & 0.453** & 0.399** & \textbf{(0.909)} &  &  &  &   &  &  &  &  &  &  &  &  &  &  &  &  &  &  &  & \\
         
         6 & 0.693** & 0.482** & 0.673** & 0.485** & 0.569** & \textbf{(0.883)} &  &  &   &  &  &  &  &  &  &  &  &  &  &  &  &  &  &  & \\
         
         7 & 0.667** & 0.626** & 0.515** & 0.447** & 0.700** & 0.689** & \textbf{(0.805)} &  &   &  &  &  &  &  &  &  &  &  &  &  &  &  &  &  & \\
         
         8 & 0.707** & 0.552** & 0.618** & 0.524** & 0.670** & 0.841** & 0.677** & \textbf{(0.867)} &   &  &  &  &  &  &  &  &  &  &  &  &  &  &  &  & \\
        
         9 & 0.481** & 0.269** & 0.528** & 0.327** & 0.352** & 0.766** & 0.347** & 0.509** & \textbf{(0.779)}  &  &  &  &  &  &  &  &  &  &  &  &  &  &  &  & \\
         
         10 & 0.138* & \textminus0.043 & 0.256** & 0.094 & \textminus0.070 & 0.547** & 0.000 & 0.214** & 0.296**  & \textbf{(0.811)} &  &  &  &  &  &  &  &  &  &  &  &  &  &  & \\
         
         11 & 0.757** & 0.767** & 0.534** & 0.512** & 0.649** & 0.563** & 0.652** & 0.636** & 0.362**  & \textminus0.017 & \textbf{(0.933)} &  &  &  &  &  &  &  &  &  &  &  &  &  & \\
        
         12 & 0.747** & 0.749** & 0.513** & 0.544** & 0.594** & 0.520** & 0.607** & 0.594** & 0.338**  & \textminus0.032 & 0.954** & \textbf{(0.963)} &  &  &  &  &  &  &  &  &  &  &  &  & \\
        
         13 & 0.741** & 0.633** & 0.605** & 0.540** & 0.586** & 0.629** & 0.584** & 0.661** & 0.436**  & 0.123 & 0.818** & 0.715** & \textbf{(0.847)} &  &  &  &  &  &  &  &  &  &  &  & \\
        
         14 & 0.333 & 0.466** & 0.193** & 0.091 & 0.449** & 0.266** & 0.426** & 0.328** &  0.130* & \textminus0.100 & 0.671** & 0.474** & 0.402** & \textbf{(0.781)} &  &  &  &  &  &  &  &  &  &  & \\
      
         15 & 0.100 & 0.031 & 0.124 & 0.081 & 0.017 & \textminus0.042 & \textminus0.066 & \textminus0.080 & 0.075  & \textminus0.054 & 0.068 & 0.102 & 0.059 & \textminus0.050 & \textbf{(0.768)} &  &  &  &  &  &  &  &  &  & \\
       
         16 & \textminus0.198** & \textminus0.125** & \textminus0.097 & \textminus0.142* & \textminus0.259** & \textminus0.181** & \textminus0.299** & \textminus0.272** & \textminus0.024  & 0.061 & \textminus0.134* & \textminus0.113 & \textminus0.178** & \textminus0.064 & 0.611** & \textbf{(0.476)} &  &  &  &  &  &  &  &  & \\
         
         17 & 0.124 & 0.070 & 0.122 & 0.112 & 0.095 & \textminus0.043 & 0.022 & \textminus0.017 &  0.037 & \textminus0.159* & 0.133* & 0.157* & 0.107 & 0.016 & 0.852** & 0.404** & \textbf{(0.620)} &  &  &  &  &  &  &  & \\
         
         18 & 0.217** & 0.161* & 0.200** & 0.154* & 0.121  & 0.072 & 0.048 & 0.038 & 0.129 & \textminus0.011 & 0.107 & 0.139* & 0.140* & \textminus0.068 & 0.817** & 0.190** & 0.540** & \textbf{(0.703)} &  &  &  &  &  &  & \\
        
         19 & \textminus0.109 & \textminus0.181** & \textminus0.006 & \textminus0.090 & \textminus0.136* & \textminus0.031 & \textminus0.121 & \textminus0.113 & 0.107  & 0.025 & \textminus0.096 & \textminus0.091 & \textminus0.088 & \textminus0.056 & 0.223** & 0.235** & 0.148* & 0.154* & \textbf{(-)} &  &  &  &  &  & \\
        
         20 & 0.310** & 0.350** & 0.179** & 0.225** & 0.392** & 0.040 & 0.329** & 0.187** & \textminus0.088  & \textminus0.276** & 0.325** & 0.292** & 0.313** & 0.219** & 0.036 & \textminus0.301** & 0.160* & 0.130* & \textminus0.130 & \textbf{(0.833)} &  &  &  &  & \\
        
         21 & 0.209** & 0.294** & 0.086 & 0.123 & 0.334** & \textminus0.137* & 0.159* & 0.012 & \textminus0.155*  & \textminus0.372** & 0.273** & 0.275** & 0.236** & 0.137* & 0.234** & \textminus0.122 & 0.289** & 0.285** & \textminus0.069 & 0.821** & \textbf{(0.778)} &  &  &  & \\
       
         22 & 0.306** & 0.294** & 0.209** & 0.248** & 0.325** & 0.188** & 0.383** & 0.288** & \textminus0.001  & \textminus0.106 & 0.273** & 0.218** & 0.287** & 0.227** & \textminus0.154* & \textminus0.371 & \textminus0.006 & \textminus0.050 & \textminus0.145* & 0.856** & 0.408** & \textbf{(0.801)} &  &  & \\
        
         23 & 0.595** & 0.591** & 0.423** & 0.414** & 0.645** & 0.569** & 0.629** & 0.632** &  0.320** & 0.065 & 0.663** & 0.660** & 0.575** & 0.346** & 0.007 & \textminus0.214** & 0.044 & 0.116 & \textminus0.117 & 0.229** & 0.158* & 0.223** & \textbf{(-)} &  & \\
        
         24 & 0.582** & 0.552** & 0.428** & 0.424** & 0.570** & 0.558** & 0.594** & 0.683** & 0.279**  & 0.054 & 0.611** & 0.572** & 0.591** & 0.350** & \textminus0.047 & \textminus0.261** & 0.017 & 0.064 & \textminus0.139* & 0.249** & 0.100 & 0.308** & 0.704** & \textbf{(-)} & \\
        
         25 & 0.148* & 0.184** & 0.124 & 0.005 & 0.224** & 0.098 & 0.187** & 0.155* & 0.100  & \textminus0.148* & 0.220** & 0.174** & 0.187** & 0.231** & \textminus0.052 & \textminus0.104 & 0.011 & \textminus0.042 & 0.024 & 0.032 & 0.054 & 0.002 & 0.192** & 0.217** & \textbf{(-)} \\

         \bottomrule
    \end{tabular}}
    \begin{flushleft}
        \textit{**Correlation is significant at the 0.01 level (two-tailed), *Correlation is significant at the 0.05 level (two-tailed). Cronbach's alpha values are in the diagonal within the brackets. 1: EmpathiSEr-P Total, 2: EmpathiSEr-P CE, 3: EmpathiSEr-P AE, 4: EmpathiSEr-P ER, 5: EQ-40, 6: IRI Total, 7: IRI PT, 8: IRI EC, 9: IRI FS, 10: IRI PD, 11: EMPA-D Total, 12: EMPA-D EIPT, 13: EMPA-D PE, 14: EMPA-D SA, 15: NPI-13 Total, 16: NPI-13 EE, 17: NPI-13 LA, 18: NPI-13 GE, 19: SINS, 20: BIDR-16 Total, 21: BIDR-16 SDE, 22: BIDR-16 IM, 23: Empathy (Personal Attribute), 24: Compassion, 25: Humour.}
    \end{flushleft}
\end{table*}

\section{Discussion}

\subsection{Revisiting the Conceptualisation of Empathy in SE} \label{sec:Revisiting the Conceptualisation of Empathy in SE}
We initially adopted Guthridge and Giummarra's empathy definition, which highlights cognitive and affective dimensions (Section \ref{sec:Conceptualisation of Empathy}) \cite{guthridge2021taxonomy}. This definition guided the early stages of scale development, including item pool generation, pretest survey, and first round of expert validation. However, all empathy experts in our first round recommended incorporating a behavioural component, emphasising the importance of observable empathic actions in professional settings (Section \ref{sec:Expert Evaluation}). This recommendation was later substantiated through the psychometric validation process, which revealed a robust three-factor structure comprising cognitive, affective, and behavioural dimensions of empathy (Section \ref{sec:EFA Results for the EmpathiSEr-P Scale}, \ref{sec:EFA Results for the EmpathiSEr-U Scale}). \textcolor{black}{Based on these converging lines of evidence, we operationalise empathy in this study as:}

\begin{boxDef}
    \begin{footnotesize}

     The ability to cognitively understand, affectively resonate with, and behaviourally respond to another person's mental and emotional states and perspectives, whilst maintaining a distinct sense of self, in order to foster mutual understanding.

    \end{footnotesize}
\end{boxDef}



\textcolor{black}{This definition builds on core psychological components of empathy and incorporates both its internal experiential dimension and its external behavioural expression, reflecting the view that empathy is not only a cognitive-affective state but also a relationally enacted phenomenon. Importantly, the definition is intended as a general conceptualisation of empathy rather than a domain-exclusive specification.} 
\textcolor{black}{Hence, it provides the theoretical foundation for the EmpathiSEr scales, which operationalise empathy through internal cognitive and affective experiences as well as their expression through empathic responses. Scores should therefore be interpreted as reflecting this multidimensional operationalisation rather than either internal experience or behavioural expression alone.}


\subsection{Implications for Research} \label{sec:Implications for Research}

\textcolor{black}{The principal value of the EmpathiSEr scales lies in the research they make possible. Although prior studies have established the relevance of empathy to SE, important questions remain about how empathy varies across roles and contexts, how it relates to practitioner, team, user, and project outcomes, and whether it can be supported through effective interventions. EmpathiSEr-P and EmpathiSEr-U provide a common measurement foundation for investigating these questions systematically. They are intended to complement qualitative and behavioural evidence rather than reduce empathy to a score or treat measurement as an end in itself.}

\noindent \faIcon{graduation-cap} \textbf{Empirical investigations of empathy:} This study introduces psychometrically validated, SE-specific empathy scales, enabling rigorous empirical and theoretical investigations in the field. The EmpathiSEr-P and EmpathiSEr-U instruments provide researchers with reliable and domain-specific tools to measure empathy in both team-based and user-facing contexts. These scales open new avenues for examining the role of empathy in areas such as software team dynamics, leadership practices, user-centred design, stakeholder engagement, and developer well-being \cite{gunatilake2025theory}. 

\noindent \faIcon{graduation-cap} \textbf{Validation through confirmatory factor analysis (CFA):} While this study employed exploratory factor analysis to establish the initial dimensional structure of scales, future research should conduct CFA to validate this factor structure in independent samples \cite{drouet2024development, hox2021confirmatory, cabrera2010author}. CFA can test the stability of the proposed dimensions and help refine the model for broader generalisability across SE contexts.

\noindent \faIcon{graduation-cap} \textbf{Intervention evaluation and longitudinal research:} The scales can be used to evaluate the effectiveness of empathy-enhancing interventions, such as empathy training or agile coaching. They also provide a basis for longitudinal studies tracking the development of empathy over time, across different career stages or organisational changes \cite{gunatilake2025manifestations}.

\noindent \faIcon{graduation-cap} \textbf{Cross-cultural and role-based studies:} Given the global and collaborative nature of SE, the scales enable cross-cultural comparisons of empathic tendencies, as well as investigations into how empathy manifests across different roles (e.g., developer, tester, UX designer, project manager). These studies could illuminate contextual factors that shape empathic behaviours in software teams \cite{gunatilake2025theory}.

\noindent \faIcon{graduation-cap} \textbf{Empathy development in SE education:} The scales can be adapted to assess and support empathy development in SE education. They enable researchers and educators to systematically evaluate how empathy evolves among students particularly during team-based activities such as capstone projects, internships, or collaborative coursework \cite{gunatilake2025theory}. Insights from such studies can inform the design of curricula, teaching strategies, and professional development programs that cultivate empathy as a core professional skill, shaping a more empathetic next generation of practitioners. 

\noindent \textcolor{black}{\faIcon{graduation-cap} \textbf{Broader applicability:} Future research can examine whether the scales are transferable to other socio-technical domains such as HCI, HRI, and AI. These areas also involve human–technology interactions in which empathy is relevant. The scales may further provide a structured basis for exploring empathic behaviours in systems that simulate human-like interaction, including conversational AI agents. However, our scales were developed and validated exclusively with SE practitioners, and their applicability beyond SE contexts remains an open empirical question.}

\subsection{\textcolor{black}{Potential Applications of the EmpathiSEr Scales}} \label{sec:Implications for Practice}

\textcolor{black}{Our EmpathiSEr scales are intended to support reflection, discussion, and research on empathy within SE contexts, like the Jefferson scale has been used to enhance empathy in medical student training \cite{hojat2016jefferson, hojat2007empathy}. However, they should not be used as standalone instruments for performance evaluation, recruitment, promotion, or other high-stakes decision making purposes. The scales do not provide diagnostic categories, nor should lower scores be interpreted as evidence of a lack of care, professional capability, or suitability for a particular role. Empathy is a dynamic and context-dependent construct that may vary across roles, teams, organisational environments, and communication settings. Hence, scores should be interpreted cautiously and alongside complementary qualitative insights. While prior research has linked empathy in SE to outcomes such as collaboration quality, communication effectiveness, practitioner well-being, and technical outcomes \cite{gunatilake2025theory, gunatilake2024enablers, cerqueira2024empathy, cerqueira2025exploring, devathasan2025empathy}, further research is needed to establish how scores obtained using the EmpathiSEr scales relate to organisational, team, and project outcomes over time.}

\textcolor{black}{The following applications are intended as illustrative examples of how the EmpathiSEr scales may be used in practice. While the scales provide a structured means of assessing empathy in SE contexts, further research is needed to examine how empathy scores relate to downstream organisational and project outcomes.}

\noindent \faIcon{laptop} \textbf{Enhancing team dynamics and collaboration:} \textcolor{black}{The EmpathiSEr-P scale can be used to assess empathy among software practitioners and may help teams reflect on patterns of communication, collaboration, and interpersonal experiences. Such insights may support discussions around team dynamics and opportunities for improvement \cite{gunatilake2025guidelines}.}

\noindent \faIcon{laptop} \textbf{Advancing user-centred development:} \textcolor{black}{The EmpathiSEr-U scale can be used to assess practitioner empathy towards users, providing organisations with a structured way to reflect on user-oriented perspectives during software development. Such insights may support user-centred development practices and discussions around user needs, preferences, and challenges \cite{gunatilake2025theory}.}

\noindent \faIcon{laptop} \textbf{Professional development and training:} \textcolor{black}{These scales may be explored in the context of training programs to support reflection on empathy-related behaviours and provide a mechanism for evaluating empathy focused training interventions such as perspective-taking \cite{gunatilake2025theory}.}

\noindent \faIcon{laptop} \textbf{Fostering empathy driven team culture and continuous improvement:} \textcolor{black}{Regular use of the scales as reflective tools may support ongoing discussions about empathy within software teams.} \textcolor{black}{Similar to assessment instruments used in software process improvement, they may help teams establish a baseline, reflect on current practices, and examine change over time. However, the scales do not define maturity levels or prescribe a standard improvement pathway. Used in this way, the scales may contribute to more respectful and inclusive team cultures and encourage continuous reflection on team interactions \cite{gunatilake2025theory, gunatilake2025guidelines}.}



\noindent \faIcon{laptop} \textbf{Remote and distributed team effectiveness:} \textcolor{black}{With the rise of remote and hybrid work models, maintaining empathic connection is increasingly important. The scales may provide a structured mechanism for reflecting on empathy-related aspects of communication and collaboration across distributed teams \cite{gunatilake2025theory}.}

\section{Limitations} \label{sec:Limitations}
While this study presents the first psychometrically validated empathy scales tailored for SE, several limitations should be acknowledged.

The responses to scale items are inherently subjective and may be interpreted differently by practitioners depending on their individual context, personality, and cultural background. Although we made considerable efforts to ensure clarity and contextual relevance, some may interpret the same response options in varied ways, introducing inconsistencies in measurement.
In addition, responses provided by practitioners may be influenced by their current work environment, team dynamics, or specific professional situations. Many practitioners reported drawing primarily from their most recent or salient workplace experiences when completing the scales, which may not fully capture their broader empathic dispositions. Further, empathy-related behaviours are often situational and relational. For instance, a respondent may behave differently when interacting with a close colleague than with distant team members or those outside their immediate group. This variability is difficult to account for in standardised scale items.

\textcolor{black}{Participants from the initial scale development survey were not explicitly excluded from participating in the subsequent EFA validation survey, hence some participants may have taken part in both phases. Where such overlap occurred, prior exposure to the study topic or earlier versions of the scale items may have influenced responses in the validation survey, for example through familiarity with the constructs or item wording.}
\textcolor{black}{The cognitive interviews involved a small and geographically concentrated sample. Although appropriate for identifying issues in item comprehension, interpretation, and response processes, the sample may not have captured the full range of interpretation issues, particularly those arising from different cultural and linguistic contexts. Further cognitive testing with participants from a broader range of cultural backgrounds may identify additional issues and support further refinement of the scales.}
While the scales were designed to be broadly applicable across roles in SE, certain items may resonate more strongly with some roles than others. For example, user-facing roles such as UX designers or product managers may find the user empathy items more directly applicable than those in purely technical or infrastructure-focused roles.
Cultural variation also plays a significant role. In some organisational or national cultures, especially those with hierarchical and authoritative norms, certain expressions of empathy, such as emotional openness or efforts to ensure that all voices are heard, may not be encouraged or may even be perceived as a weakness. This may affect how participants from different backgrounds respond to specific items.

The complexity of human behaviour presents another challenge. Although the scales aim to capture key dimensions of empathic cognition, affect, and behaviour, they inevitably simplify a deeply nuanced and dynamic phenomenon. Some participants noted that empathy in practice is not always emotionally expressive and may involve boundaries or restraint. Others expressed concern that certain scale items could oversimplify the subtleties of real-world empathic engagement. These instruments should therefore be used as a tool to guide reflection and assessment, rather than as definitive measures of empathic ability.
\textcolor{black}{As with many socio-technical and human-centred constructs, quantitative empathy scales may not fully capture the complexity and situated nature of empathic experiences in practice. Hence, the proposed instruments are intended to complement, rather than replace, qualitative and mixed-method approaches that can provide richer contextual insight into practitioners' lived experiences and interactions.}
\textcolor{black}{The proposed scales assess self-reported expressions of empathy within SE contexts. However, empathy may be experienced and expressed differently across neurodivergent practitioners, whose communication styles and interpersonal behaviours may not always align with dominant workplace norms. Consequently, some scale items may capture conventionally recognised or neurotypical expressions of empathy rather than the full diversity of empathic experiences. Future work should therefore examine the applicability, interpretability, and inclusiveness of the scales across neurodiverse software practitioners.}
\textcolor{black}{Further, as with all self-report measures, responses may be influenced by social desirability bias and individual differences in self-reflective ability, which should be considered when interpreting empathy scores in practice.}

The ER subscale of EmpathiSEr-U included one item that exhibited a small cross-loading on another factor, indicating potential ambiguity in its interpretation. Although retained for conceptual and statistical reasons, this item may affect the scale's factorial clarity. Further validation in larger and more diverse samples is needed to examine its performance and explore possible revision. 
Finally, while exploratory factor analysis was used to establish the initial factor structure of the scales, further validation is needed. Future research should apply CFA to test the dimensional stability of the instruments in independent samples. Additional work is also needed to assess the scales' cross-cultural robustness and their sensitivity to change over time, particularly in response to empathy training or team interventions.

\section{Conclusion} \label{sec:Conclusion}
Empathy is increasingly recognised as a vital human aspect in SE, shaping collaboration, user engagement, and decision-making. 
\textcolor{black}{However, further research is needed to understand how empathy operates across SE roles and contexts, how it relates to outcomes, and how empathy-related practices can be effectively supported. A major obstacle to such research has been the absence of reliable instruments that reflect the distinctive socio-technical contexts of SE.}
This study addressed that gap by developing and validating two domain-specific instruments: the EmpathiSEr-P scale, which measures empathy among software practitioners, and the EmpathiSEr-U scale, which assesses practitioner empathy towards users.
The scale development followed a rigorous, multi-phase methodology that integrated established psychological measures with empirical insights from the SE domain. The items were refined through expert evaluation, cognitive interviews, and statistical analysis, resulting in two psychometrically sound instruments that capture cognitive, affective, and behavioural dimensions of empathy.
\textcolor{black}{These scales offer a measurement foundation for both researchers and practitioners for systematically investigating the role of empathy in SE and evaluating initiatives intended to support empathy-related practices.} They support more systematic investigations into the role of empathy across roles and workflows, and facilitate the evaluation of empathy-enhancing interventions. 
This work lays important groundwork for integrating empathy more deeply into SE research, teaching, and everyday practice.

\section*{Acknowledgments}
Gunatilake and Grundy are supported by ARC Laureate Fellowship FL190100035. We sincerely thank all participants for generously sharing their experiences and insights. 
This work was also supported by the Monash Statistical Consulting Service (Dr Tim Powers), part of Monash eResearch.


\bibliographystyle{ieeetr}
\bibliography{References}

\end{document}